# A novel general modeling of the viscoelastic properties of fluids: application to mechanical relaxation and low frequency oscillation measurements of liquid water

F. Aitken and F. Volino

Univ. Grenoble Alpes, CNRS, Grenoble INP, G2ELab, F-38000 Grenoble, France.

**Abstract**. The aim of this paper is to calculate the time dependence of the mean position (and orientation) of a fluid particle when a fluid system at thermodynamic equilibrium is submitted to a mechanical action. The starting point of this novel theoretical approach is the introduction of a mechanical energy functional. Then using the notions of inertial modes and action temperature, and assuming a mechanical energy equipartition principle per mode, the model predict the existence of a dynamic phase transition where the rheological behavior of the medium evolves from a solid-like to a liquid-like regime when the mechanical action is increased. The well-known Newtonian behavior is recovered as limiting case. The present modeling is applied to the analysis of recent liquid water viscoelastic data pointing out a prevalent elastic behavior in confined geometry. It is demonstrated that the model makes it possible to understand these data in a coherent and unified way with the transport properties (viscosity and self-diffusion coefficient). It is concluded that any finite volume of fluid at rest possesses a static shear elasticity and should therefore be considered as a solid-like medium.

## 1 Introduction

The solid-like behavior of many liquids at low frequencies has been experimentally demonstrated by many authors (e.g. Ref. 1 to 6). However, this property of liquids is difficult to reconcile with the various classical theoretical approaches (e.g. Ref. 7 to 9). Only recently, some work has addressed the problem of theoretical explanation in the framework of models based on standard statistical physics (e.g. see Ref. 10), and a scaling law has been proposed to account for some of these experimental results (Ref. 10a).

In this paper, we will develop a groundbreaking theoretical model having no equivalent to our knowledge, such that it will lead us to admit that any finite volume of a liquid at rest must be considered as a solid with low but finite static shear elasticity. To do this, we will strongly rely on the space domain modeling developed in Ref. 11 and will extend it to the time domain. Indeed, in Ref. 11 a non-conventional lattice model has been developed to describe the thermodynamic equilibrium properties of a fluid medium. In this model, the cell of the lattice, called "basic unit", is composed *a priori* of several molecules or atoms. Due to the temperature, the center of mass of this basic unit fluctuates around its mean position $\vec{r}_f$.

The departure $\vec{u} = \vec{r} - \vec{r}_f$ of the position $\vec{r}$ from this mean position $\vec{r}_f$ is assumed to be a Gaussian random variable of the variance $\langle u^2 \rangle$. The quantities $\vec{r}_f$ and $\sqrt{\langle u^2 \rangle}$ might thus be identified as the center and the size of the "thermal cloud", respectively, in which each basic unit is delocalized. The aim of this model was essentially to calculate the variance $\left\langle u^2 \right\rangle$, and to establish expressions for the self-diffusion coefficient and the viscosity in terms of this variance and time scales phenomenologically introduced. It was shown that, for water, this model can reproduce quantitatively, in the whole phase diagram, a very large amount of self-diffusion coefficient data, and of viscosity data in the context of Newtonian fluids. The model implies that the Newtonian behavior is a part of a much wider rheological behavior. This model, developed in the space domain, was limited to the calculation of the thermodynamic equilibrium, which implies that the mean position $\vec{r}_f(t)$ of the Gaussian distribution of $\vec{u}$ is assumed to be static. In other words, $\vec{r}_f(t) = \vec{r}_f(0) = \vec{r}_f$ at all times. The extension of the model to the time domain consists in calculating the evolution of $\vec{r}_f(t)$ when the system is pushed out-of-equilibrium via an external mechanical perturbation. The mean position $\vec{r}_f(t)$ can be identified as the "point" in fluid mechanics at which the macroscopic mechanical actions are applied. This makes it possible to define the notion of "fluid particle" as the volume of the thermal cloud and whose centre of gravity is $\vec{r}_f(t)$. Therefore, the present purpose is to describe rheological properties of a fluid in the framework of this lattice-like model. In particular, this development will allow us to justify the phenomenological relations introduced in Ref. 11 to describe transport coefficients.

The fundamental ingredient of the model is the static shear elasticity parameter $K$. Despite this concept has allowed us to analyze self-consistently the very large number of transport coefficients data of water, potassium and thallium (Refs. 11 and 12), the evidence that a static shear elasticity exists in fluids was only indirect: it is simply a parameter of the model. To show that this shear elasticity is associated with a measurable quantity, it is very important to analyze experimental data which can only be analyzed in a standard way by the necessary existence of a static shear elasticity. Today such data exist, obtained from mechanical relaxation and very low frequency oscillatory measurements (e.g. Ref. 13), especially for liquid water (Ref. 14). They are extremely valuable for the present purpose. As an application of the model, we present and analyze them using two different approaches: firstly, the relaxation data are analyzed in the light of the present modeling, and second, the dynamic viscoelastic data are analyzed with the classical approach commonly used in the field of rheology. It is shown that the values of the parameters determined by the two methods are consistent with each other, in particular for the elastic shear modulus. In addition, the analysis is extended to the shear elasticity data of the literature obtained in the KHz and MHz ranges.

The present modeling is applied here to liquid water data only (in spite of the fact that experimental data are available for other fluids) because the model parameters are the same as those used to describe viscosity and self-diffusion coefficient (i.e. these parameters are known from Ref. 11) and because of the availability of large and varied rheological data sets for this fluid.

## 2 The general basis of the modeling

In this section, we focus on the dynamic aspects which result from a mechanical action made on a "fluid" system initially at thermodynamic equilibrium. This action, which is carried out with the use of macroscopic forces or torques, induces a "net" translational and/or rotational motion of the fluid particles. The aim here is to calculate the position and orientation of the fluid particles, as a function of the nature and intensity of the mechanical action, and of the fluid characteristics, within the framework of a non-extensive theory "in duration". By analogy with the elastic energy functional of the "volume" theory (Ref. 11), the starting point of this "duration" theory is a mechanical energy functional associated with the mechanical action. We will see that this operating way makes it possible to elaborate a simple formalism, where dissipation and irreversible time aspects appear as fundamental ingredients of the model.

### 2.1. Mechanical energy functional

Because the action (in the mechanical sense, i.e. a mechanical energy supplied for a given time) of forces or torques moves the system out-of-thermodynamic equilibrium, the starting point of this duration theory is consequently a mechanical energy functional (i.e. representing an action per unit time), which will be associated with each point $X$ of the fluid. As a result of the action, a point of the fluid $X$ undergoes a translational relative displacement $x_f(X, t_1)$ and/or an orientational relative displacement $\varphi_f(X, t_1)$, with respect to a reference frame linked to a fixed boundary of the system at a time $t_1$.

Inspired by the fundamental law of dynamics, we assume that, to each particle of fluid located at point $X$, we can associate a mechanical energy functional $F_A$ such that:

- for the translation: 
$$F_A^{\mathbf{t}} = c_0 \int_{\substack{t = \text{duration} \\ \text{of action}}} K_A^{\mathbf{t}} \, \gamma_f^{\mathbf{t}}(X, t_1) \, dt_1 \tag{1}$$

- for the rotation: 
$$F_A^{\mathbf{r}} = \Omega_0 \int_{\substack{t = \text{duration} \\ \text{of action}}} K_A^{\mathbf{r}} \, \gamma_f^{\mathbf{r}}(X, t_1) \, dt_1 \tag{2}$$

where $c_0$ and $\Omega_0$ are characteristic velocities (more precisely an angular velocity in the second case) with which the mechanical information is propagated in the medium.

The index A is used for concepts related to the action, and the superscripts "$\mathbf{t}$" or "$\mathbf{r}$" to translation or rotation, respectively. Thus, $\gamma_f^{\mathbf{t}}(X, t_1)$ and $\gamma_f^{\mathbf{r}}(X, t_1)$ are the translational and rotational accelerations at time $t_1$, at the point $X$ of the fluid considered and $K_A^{\mathbf{t}}$ and $K_A^{\mathbf{r}}$ are quantities which have the dimensions of a mass and a moment of inertia, respectively.

Unlike standard classical mechanics where kinematics quantities at time $t_1$ (i.e. position, velocity, acceleration), are analytical functions of the single variable $t_1$, in the present model, these quantities are defined by two times $t_1$ and $t_2$, possibly as close to each other ($\Delta t_{12} = t_1 - t_2$). This implies that kinematics quantities are functions of two (independent) time variables rather than one. The corresponding standard quantities are therefore only the limits of these functions when the values of the two variables become equal. For example, the translational acceleration at time $t_1$, along the direction $x$, must be written as:

$$\gamma_f^{\mathbf{t}}(X, t_1) = \lim_{t_2 \to t_1} \left( \frac{\partial^2 x_f(X, t_1, t_2)}{\partial t_1 \partial t_2} \right) \tag{3}$$

and the velocity:

$$v_f^{\mathbf{t}}(X, t_1) = \frac{1}{2} \lim_{t_2 \to t_1} \left( \frac{\partial x_f(X, t_1, t_2)}{\partial t_1} + \frac{\partial x_f(X, t_1, t_2)}{\partial t_2} \right) \tag{4}$$

The necessity to introduce two close instants $t_1$ and $t_2$ to define the kinematics quantities means that, for the experimental physics, the description of the time dependence of a (dissipative) phenomenon is not strictly continuous and requires a minimum time delay $t_2 - t_1 = \tau_{\min}$. In the present particular case where we describe the collective displacement of material objects with a finite size (here the basic units), consecutive to the application of an external perturbation, this time delay is logically identified with the time required for the shear information to propagate across the objects, namely $\tau_{\min} = l_{\text{obj}}/c_0$ where $l_{\text{obj}}$ is the size of the objects. For usual liquids, $\tau_{\min}$ is of the order of a fraction of picosecond. Since in the description of usual flow phenomena, the times involved are generally (much) longer, $\tau_{\min}$ can be neglected and, in practice, the kinematics quantities can be considered as continuous functions of time.

On the basis of these statements, we decompose $x_f(X, t_1, t_2)$ into double temporal Fourier series on the duration $t$. The notion of independent **inertial modes** (inertial because characterized by a (inertial) mass or a moment of inertia) is then introduced, each of them being characterized by a pulsation $\omega$. The values of these pulsations range between a maximum value $\omega_c$, which is a characteristic of the system (which must be described as "dynamic" since the entire description only makes sense if there is net fluid motion), and a minimum value $\omega_c/N_A$ where the index $N_A$ is the "temporal" equivalent of the "spatial" index $N$ introduced in Ref. 11 to describe elastic modes. We get:

$$x_f(X, t_1, t_2) = \sum_{\omega, \omega'} x_{\omega, \omega'} \exp(i\omega t_1) \exp(i\omega' t_2) \qquad (5)$$

where $x_{\omega, \omega'}$ are the Fourier coefficients.

By analogy with the wave vectors of elastic modes defined in a volume, we assume that the pulsations of the inertial modes are defined in a duration $t$. It is then deduced in a similar way that the density of inertial modes $\rho_A(\omega)$ (number of inertial modes per unit pulsation) in the duration $t$ is given by:

$$\rho_A(\omega) = t \qquad (6)$$

by analogy with $\rho_E(\vec{q}) = 3V/(2\pi)^3$ for the spatial case. Here, there is not the factor 3 because duration is a "one-dimensional space". The absence of a $2\pi$ factor in Eq. (6) is taken as a convention. The index:

$$N_A = \omega_c\, t \qquad (7)$$

represents the number of "elementary" $\omega_c^{-1}$ durations in duration $t$. It will be called the **reduced duration**, by analogy with the reduced size $N$ for elastic modes. Now the number of inertial modes $\mathfrak{N}_A$ for the duration $t$ associated with the fluid point under consideration is obtained by summing $\rho_A(\omega)$ between $\pm\, \omega_c/N_A$ and $\pm\, \omega_c$ (the $\pm$ sign comes from the fact that the pulsations can be positive or negative). We get:

$$\mathfrak{N}_A = 2(N_A - 1) \qquad (8)$$

Since $\mathfrak{N}_A$ is a positive number, Eq. (8) shows that as long as the duration is less than $\omega_c^{-1}$, this description is not applicable. The movement between $t = 0$ and $t = \omega_c^{-1}$ will be identified as a **transient regime**.

By carrying Eq. (5) into Eq. (3), then Eq. (3) into Eq. (1) and taking into account the orthogonality of complex exponential functions:

$$\int_0^t \exp\big(i(\omega + \omega')t_1\big)dt_1 = t\,\delta_{\omega,-\omega'}$$

where $\delta_{\omega,-\omega'}$ is the Kronecker symbol, Eq. (1) can now be simply written:

$$F_A^t = c_0 t \sum_\omega \big(K_A^t \omega^2 x_{\omega,-\omega}\big) \qquad (9)$$

The same approach for the rotation leads to replace Eq. (2) by the following relationship:

$$F_A^r = \Omega_0 t \sum_\omega \big(K_A^r \omega^2 \varphi_{\omega,-\omega}\big) \qquad (10)$$

Contrary to thermal motions where it has been shown in Ref. 11 that it is the mean square of the displacements that are added, here for mechanical motions, it is the displacements (or angular displacements) that are added in the expression of the energy functional. Another fundamental difference is that the "volume of space" defined by $N$ is in principle fixed while the "volume of time" defined by $N_A$ increases, possibly indefinitely.

In this analysis, the effects of the stress on the displacements inside the basic units, producing a deformation of these objects, have not been taken into account, i.e. only "viscous" aspects associated with long-distance displacements are considered. These short non-dissipative distance displacements exist but are neglected in this present analysis. To take into account such short time effect, a feedback term which introduces a constant stiffness must be added in the mechanical energy functionals Eqs. (1) and (2). In these more complex cases, the full mechanical displacement will be the sum of an elastic displacement representing the reversible deformation of the basic units and the irreversible viscous displacement described by the present modeling. Such more complex modeling will be presented elsewhere.

## 2.2. Concept of action temperature and equipartition principle of the mechanical energy

To go further, we introduce the notion of "action temperature", noted as $T_A$, that we will associate to each point $X$ of the fluid. This action temperature is a dimensionless number that describes the strength of mechanical action made on the system at the point $X$ of the fluid. If the action is independent of duration, it can therefore be stated that, during this action, the fluid point is at operating temperature $T_A$. For a fluid under flow, the operating action temperature $T_A$ is zero at any point where the velocity is zero, which is generally the case on fixed walls (see below). For a fluid at thermodynamic equilibrium, all fluid points are at zero action temperature.

The natural unity for action being the Planck constant $\hbar$, any action may be written as $\hbar T_A$. By transposing here to the mechanical energy the equipartition principle of energy relevant to the space domain, we postulate an **equipartition principle of the mechanical energy** ***associated with the action***, namely that the mechanical energy per inertial mode, for a fluid point $X$ at action temperature $T_A$, is $\hbar\omega_c T_A/2$. Note that if we had defined an "action functional" instead of a "mechanical energy functional", this equipartition principle would be such that the action by inertial mode is equal to $\hbar T_A/2$. The two descriptions are equivalent.

By explicitly writing this equipartition principle, we obtain the expression of the contribution $x_{\omega,-\omega}$ (more precisely written $x_{f,\omega}$) of the translational inertial mode of pulsation $\omega$ to the displacement $x_f$, (or of the contribution $\varphi_{\omega,-\omega}$ of the rotational inertial mode $\omega$ to the rotation $\varphi_f$) of the fluid at the point $X$ considered for the duration $t$. We have:

- for the translation: $x_{f,\omega}(X,t) = \dfrac{1}{2}\dfrac{\hbar\omega_c T_A}{c_0\ K_A^{\mathbf{t}}\ \omega^2 t}$          (11)

- for the rotation: $\varphi_{f,\omega}(X,t) = \dfrac{1}{2}\dfrac{\hbar\omega_c T_A}{\Omega_0\ K_A^{\mathbf{r}}\ \omega^2 t}$          (12)

The next step is to define, as for the spatial case, coefficients $K_A$ not as usual mass or inertia moments, but as quantities associated with each of the inertial modes of pulsation $\omega$, functions of $\omega$ according to a power law with an exponent $\nu_A$ that depends on the action temperature $T_A$, the same in both cases, such that:

$$K_A^{\mathbf{t,r}} = K_{A0}^{\mathbf{t,r}}\left|\frac{\omega}{\omega_c}\right|^{\nu_A-2}$$          (13)

where $K_{A0}^{\mathbf{t,r}}$ are a mass and inertia moment characteristic of the dynamic system, associated with dissipative phenomena. As for the case of the volume theory, it is shown in Appendix A that this assumption is mathematically equivalent of using fractional derivatives in the definition of the mechanical energy functional form.

Finally, the displacement (beyond time $\omega_c^{-1}$), caused by all the inertial modes for the duration $t$ is obtained by summing over all these modes (it is twice the sum over the positive values of $\omega$), so that:

- for the translation: $x_f(X,t) = \dfrac{\hbar T_A(X)}{c_0\ K_{A0}^{\mathbf{t}}}H_{N_A}(\nu_A)$          (14)

- for the rotation: $\varphi_f(X,t) = \dfrac{\hbar T_A(X)}{\Omega_0\ K_{A0}^{\mathbf{r}}}H_{N_A}(\nu_A)$          (15)

where $H_{N_A}(\nu_A)$ is the same function as in the spatial case that can be written:

$$H_{N_A}(\nu_A) = \frac{N_A^{\nu_A-1}-1}{\nu_A-1} = \frac{(\omega_c t)^{\nu_A-1}-1}{\nu_A-1}$$          (16)

Eq. (14) and Eq. (15) represent the very general one-dimensional solution of the displacement and angular displacement of the fluid point $X$ under external mechanical action for the duration $t$. Despite the concept of action temperature is not yet explicitly related to a familiar concept, it is clear that the stronger the action (i.e. the greater the force or torque applied at the point of the fluid, the higher the action temperature $T_A$, which results, at a given time, in a larger linear displacement $x_f$ (or a greater angular displacement $\varphi_f$).

Eq. (16) shows that, if $v_A$ increases, then the crossing by $v_A = 1$ corresponds to a very large increase of the displacement. The condition $v_A = 1$ therefore corresponds to a "dynamic" phase transition (i.e. probably a plastic transition; i.e. solid to liquid) in the same way as in the spatial case $v = 1$ corresponds to a "thermal" phase transition (i.e. probably a glass-like transition). Thus, by analogy with the spatial case, we postulate that:

$$1 - v_A = \left( \frac{T_{A0}}{T_A} - 1 \right)^{\frac{1}{2}} \quad \text{for} \quad T_A \leq T_{A0} \quad \text{(i.e. solid-like regime)} \tag{17a}$$

$$v_A - 1 = \left( 1 - \frac{T_{A0}}{T_A} \right)^{\frac{1}{4}} \quad \text{for} \quad T_A \geq T_{A0} \quad \text{(i.e. liquid-like regime)} \tag{17b}$$

where $T_{A0}$ represents a transition action temperature which is a characteristic of the dynamic system. It therefore appears that the parameter $v_A$ will vary continuously from $-\infty$ to 2 for which $H_{N_A}(-\infty) = 0$ and $H_{N_A}(2) = N_A - 1 = (\omega_c t) - 1$.

In other words, the notions of (temporal) "inertial modes" and "action temperature", and of a principle of "equipartition of the mechanical energy associated with the action", are analogous to the elastic (spatial) modes, the usual temperature and the thermal energy equipartition principle in the volume theory, this inertial mode theory predicts a "dynamic order-disorder" phase transition, the parameter governing this transition being the action temperature. Table 1 highlights the analogy of the parameters between the two theoretical models.

| *Volume theory* (elastic modes) Describes the random fluctuations of the basic units about an equilibrium position | *Duration theory* (inertial modes) Describes the deterministic motion of the equilibrium position in response to external mechanical action. |
|---|---|
| - For translation: random displacement of the center of mass: $\vec{u}$  <br><br> - For rotation: random angular displacement of the orientation: $\vec{\Omega}$ | - For translation: deterministic displacement of the mean position of a basic unit: $x_f$  <br><br> - For rotation: deterministic angular displacement of the mean orientation around a fixed axis: $\varphi_f$ |
| Elastic energy functional: $F$ | Mechanical energy functional: $F_A$ |
| Volume of the system: $V$ | Duration of the mechanical action: $t$ |
| Spatial wave-vector modulus: $q$ | Pulsation: $\omega$ |
| Density of elastic modes in volume $V$: $\rho_E(\vec{q})$ | Density of inertial modes in duration $t$: $\rho_A(\omega)$ |
| Cutoff wave-vector modulus: $q_c$ | Cutoff pulsation modulus: $\omega_c$ |
| - For translation: shear elastic constant $K$ (unit: force per unit surface)  <br><br> - For rotation: rotational elastic constant $K_r$ (unit: torque per unit length) | - For translation: mass $K_{A0}^{t}$  <br><br> - For rotation: moment of inertia $K_{A0}^{r}$ |
| Reduced fluctuative distance: $N$ | Reduced duration: $N_A$ |
| Exponent: $v$ | Exponent: $v_A$ |
| Thermodynamic temperature: $T$ | Action temperature: $T_A$ |
| Ordered-disordered transition temperature: $T_t$ | Solid-liquid transition action temperature: $T_{A0}$ |

Table 1. Equivalence table: each line indicates analogous parameters in the volume and duration theories.

Table 1 indicates that the duration $t$ is the analogue of the volume $V$ (and *vice versa*). These two parameters are fundamental for the two theoretical models. The significance of these two quantities should thus be emphasized:

- the volume $V$ represents a "piece of space" in which matter is located by the position $\vec{r}$ of its constitutive material elements (atoms, molecules, basic units, …);
- the duration $t$ represents a "piece of time" associated with a mechanical stress which is applied on the constitutive elements, producing a net displacement of the latter. The instant $t_1$ represents the analogue of the spatial position in the sense that it defines a "temporal position" in the duration $t$.

Although there is an analogy between $V$ and $t$ in the two theories, the two concepts are fundamentally different. Indeed, in the volume theory, the volume $V$ is fixed and consequently the density of elastic modes per unit wave-vector $\rho_E(\vec{q}) \propto V$ is a constant while in the duration theory, the time "flows" and consequently the density of inertial modes per unit pulsation $\rho_A(\omega) = t$ changes continuously. Let us insist again on the fact that the concept of time (in the sense of duration) is irrelevant for systems at thermodynamic equilibrium. It only emerges in a real experiment, when energy is transferred from one part of a system to another, in the form of matter movement, and it disappears when all the transferred energy has been dissipated (transformed into random motions) so that the whole system has returned to thermodynamic equilibrium. Thus the concept of "reversible time" is foreign to the present modeling.

## 2.3. Relationship closure for the translational case

In the following, we will focus only on the translational case. To close the formalism, we need to define the following parameters: $c_0$, $\omega_c$, $K_{A0}^t$, $T_A$ and $T_{A0}$.

It was pointed out in the spatial case (see Ref. 11) that the only characteristic celerity is defined by the static shear elastic constant $K$ of the medium such that:

$$c_0 = \sqrt{\frac{K}{\rho}} \tag{18}$$

where $\rho$ represents the medium mass per unit volume. (Note that for the rotational case, we have $\Omega_0 = \sqrt{K_r / J}$ where $J$ is the moment of inertia per unit length.)

The quantity $\omega_c^{-1}$ is a characteristic parameter that represents the time taken for the "global" information introduced by the mechanical action to propagate over the dissipative distance $d$. This time thus corresponds to a delay between the action and the reaction of the system to this action, during which internal phenomena occur in the system but they are not described by the present theory. It is therefore logical, in the context of the present description, to scale $\omega_c^{-1}$ with this "macroscopic" characteristic time $\tau = d/c_0$. It is also clear that the reaction delay depends on the position $X$ of the fluid point under consideration. The simplest way to express these relationships is as follows:

$$\omega_c^{-1} = \tau / f_{\omega_c}(X, \xi) \tag{19}$$

where $f_{\omega_c}(X, \xi)$ is a function which depends on the position $X$ of the object and on both the properties of the system and the overall mechanical action applied. These properties are grouped together in the single parameter $\xi$. Eq. (19) replaces an unknown with a new unknown non-dimensional function but whose physical meaning can be more easily understood through the different experiments.

In Eq. (14), $K_{A0}^t$ has the dimension of a mass that must be associated with dissipative phenomena. In Ref. 11, it has been shown that during a shearing experiment the occurrence of a released gas implies an additional dissipation which must be combined with that of the sheared liquid. Given the expression of these dissipative terms in Ref. 11, it appears that the expression of $K_{A0}^t$ should be written as follows:

$$K_{A0}^{t} = m_B \left( \frac{1}{H_N(v)} + \frac{\rho_{Knu}}{\rho} \frac{\delta}{d} \frac{\sqrt{R_g T/M}}{c_0} \right) \qquad (20)$$

where $m_B$ represents the mass of the moving basic unit and $H_N(v) = \frac{N^{v-1}-1}{v-1}$ is the function introduced to describe elastic modes. The second term represents the dissipative effect of the gas released during shearing, which corresponds here to a reaction to the mechanical action. Let's remember that $\rho_{Knu}$ is the density of this released gas, $\delta$ is a distance defined by the experimental set-up, $M$ is the molar mass of the medium and $R_g$ represents the perfect gas constant.

From the expression of Eq. (14), it is possible to identify $T_{A0}$ such that:

$$\hbar T_{A0} = c_0 K_{A0}^{t} \, l \qquad (21)$$

where $l$ is a characteristic length of the system. The quantity $\hbar T_{A0}$ can be understood as a "reaction" of the dynamic system to counteract the action made on it to set it in motion. This reaction is independent of the particular point $X$ of the fluid, and can be considered as an overall quantity that characterizes the sample as a whole.

To express the local action, it is necessary to introduce a characteristic time of this action and a characteristic energy corresponding to this operating time. The only characteristic time linked to the action is $\omega_c^{-1}$ therefore we assume that the local action can always be written in the following form:

$$\hbar T_A(X) = E_A(X) \, \mho_B \, \omega_c^{-1} \qquad (22)$$

where $\mho_B = m_B/\rho$ represents the volume per basic unit and $E_A(X)$ represents the mechanical energy per unit volume in the fluid at point $X$. Unlike $\hbar T_{A0}$ the action $\hbar T_A(X)$ is a local quantity which depends on the particular point $X$ of the fluid.

By replacing the different parameters in the above expressions, a new general expression of the translational relative displacement is deduced:

$$\left. \frac{x_f(X,t)}{l} \right|_{\text{steady}} = \frac{\hbar T_A(X)}{\hbar T_{A0}} H_{N_A}(v_A) = \frac{E_A(X)}{(K_N + K_{gas}) f_{\omega_c}(X,\xi)} \frac{d}{l} H_{N_A}(v_A) \qquad (23)$$

where $K_N = K/H_N(v)$ represents a shear elastic modulus associated with the liquid part and $K_{gas} = \rho_{Knu} \frac{\delta}{d} \sqrt{\frac{R_g T}{M}} c_0$ is a shear elastic modulus associated with the released gas. Since $\hbar T_A(X)$ depends on the particular point $X$ of the fluid, it can be deduced from Eq. (23) that there is a threshold of the mechanical energy per unit volume $E_{A,\text{threshold}}$ for which $T_A = T_{A0}$ such that:

$$E_{A,\,threshold}\left(X\right) = \frac{l}{d}\left(K_N + K_{gas}\right)f_{\omega_c}\left(X,\xi\right) \tag{24}$$

This relationship can be used to determine the point $X$ of the fluid where the dynamic phase transition occurs. This makes it possible to define a notion of "boundary layer": the points of the fluid located below the energy threshold will have a zero velocity since for all these points the displacement at "infinite" time is finite.

In the particular case where the mechanical action is constant over time, a simple formula can be deduced for the steady state velocity at point $X$:

$$v_f\left(X,t\right)\big|_{steady} = \frac{E_A\left(X\right)}{\left(K_N + K_{gas}\right)\tau}d\left(\omega_c t\right)^{v_A - 2} \tag{25}$$

It is clear from this particular case that the velocity function is a power law of time, which tends towards a constant when $v_A \to 2$.

## 2.4. The Newtonian limit

Within the general framework of the present modeling, the laminar Newtonian regime is obtained as the asymptotic limit of the liquid regime when $T_A \gg T_{A0}$ (i.e. $v_A \to 2$) and $t \gg \omega_c^{-1}$. For this limit, it is deduced:

$$\frac{x_f\left(X,t\right)}{d}\bigg|_{\substack{Newtonian \\ regime}} = \frac{E_A\left(X\right)}{\left(K_N\tau + \rho_{Knu}\delta\sqrt{\dfrac{R_g T}{M}}\right)}t \tag{26}$$

Hence,

$$\frac{v_f\left(X,t\right)}{d}\bigg|_{\substack{Newtonian \\ regime}} = \frac{E_A\left(X\right)}{\left(K_N\tau + \rho_{Knu}\delta\sqrt{\dfrac{R_g T}{M}}\right)} \tag{27}$$

From a dimensional point of view, Eq. (27) relates a velocity gradient to a shear stress by a proportionality coefficient which is characteristic of Newtonian behavior. This proportionality constant is identified as the dynamic viscosity $\eta$. Terms defining the dynamic viscosity have been postulated without justification in Ref. 11 and are therefore now justified. In addition, it has been shown that this viscosity modeling makes it possible to reproduce and understand the full range of water viscosity data. Eq. (27) can be identified with the solutions of the Navier-Stokes equations in the low Reynolds number laminar steady state regime. The solutions of the Navier-Stokes equations being largely experimentally verified in the Newtonian limit for usual viscosity experiments, it is therefore necessary that the present modeling must be in line with these solutions. This condition provides a possible means of determining the expression of the term $E_A(X)$. Indeed, since Eq. (27) is verifying the Navier-Stokes equations, then $E_A(X)$ must also verify this one with the boundary conditions deduced from the velocity boundary conditions.

Table 2 gives some examples of expressions of $E_A(X)$ for simple one-dimensional laminar shear flows for which it is possible to make a direct identification. It can be checked that $E_A(X)$ is maximum where the velocity is maximum and zero where the velocity is zero.

| | |
|---|---|
| **(a).** Simple shear conditions: flow between two parallel plates occupying the planes $z = 0$ and $z = e$. The first plate is fixed, while the second slides in the O$x$ direction with velocity $U$. One finds: $$E_A(z) = \frac{\eta U}{d}\frac{z}{e}.$$ In such case, the dissipative distance $d$ is typically $d = e$. The normal stress to the O$xy$ plane is $\sigma = \frac{\eta U}{e}$ from which one can still write: $$E_A(z) = \sigma\frac{z}{e}.$$ | **(b).** Cylinder pipe geometry: flow in a cylindrical pipe of radius $r = R$. The flow is due to a constant pressure gradient $G$ along the corresponding O$x$ direction with the cylinder axis. One finds: $$E_A(r) = \frac{GR^2}{4d}\left(1 - \left(\frac{r}{R}\right)^2\right).$$ For small capillary tubes, the dissipative distance $d$ is typically $d = R$. Knowing that the stress on the pipe wall is $\sigma = \frac{GR}{2}$, one can still write: $$E_A(r) = \frac{\sigma}{2}\left(1 - \left(\frac{r}{R}\right)^2\right).$$ |
| **(c).** Two coaxial cylinders flow: the inner cylinder is fixed and has a radius $R_1$. The outer cylinder is rotating on the common axis O$z$ of the two cylinders with an angular velocity $\Omega_2$. The outer cylinder has a radius $R_2$. One finds: $$E_A(r) = \frac{\eta\Omega_2}{d}\frac{R_2^2}{R_2^2 - R_1^2}\frac{r^2 - R_1^2}{r}.$$ If $R_1/R_2 \geq 0.99$, the dissipative distance $d$ is typically $d = R_2 - R_1$. Knowing that the torque per unit length on the moving surface is $\Gamma = \frac{4\pi\eta\Omega_2 R_2^2 R_1^2}{R_2^2 - R_1^2}$, one can still write: $$E_A(r) = \frac{\Gamma L}{4\pi R_1^2\left(R_2 - R_1\right)}\frac{r^2 - R_1^2}{r}$$ where $L$ represents the height of the cylinders. | **(d).** Disk-like plate rotating geometry: flow between two parallel planar disks occupying the planes $z = 0$ and $z = e$. The first disk is fixed, while the second rotate along the common O$z$ axis direction with an angular velocity $\Omega$. The two planar disks have the same radius $R$. One finds: $$E_A(r,z) = \frac{\eta\Omega}{d}r\frac{z}{e}.$$ In such case, the dissipative distance $d$ is typically $d = e$. Knowing that the torque on the moving surface is $\Gamma = \frac{\eta\pi\Omega R^4}{2e}$, one can still write: $$E_A(r,z) = \frac{2\Gamma}{\pi R^3}\frac{r}{R}\frac{z}{e}.$$ |

Table 2. Expression of $E_A(X)$ for four simple one-dimensional viscous incompressible flow examples.

## 2.5. The transient regime

In the foregoing, it has been emphasized that the formalism developed can only be applied to long times, i.e. for $t > \omega_c^{-1}$. In a real experiment, the motion starts as soon as the action is applied at $t = 0$. The aim is to describe the motion between 0 and $\omega_c^{-1}$. In this time interval, since no dissipation is possible, the energy supplied by the external forces can only be stored in the system. When the applied stress is constant and the initial velocity is zero, the movement at $t = 0$ is necessarily accelerated. In other words, assuming that in a short time such that $t << \omega_c^{-1}$ the system verifies the laws of classical mechanics, it can be deduced that the initial velocity must vary linearly with time and displacement as the square of time.

Assuming that the kinematics quantities (velocity, acceleration) are continuous as they cross $t = \omega_c^{-1}$, we can define different functions that have the desired properties. A simple function that gives a connection closest to $\omega_c^{-1}$ is the following one:

$$v_f(X,t) = v_f(X,t)\big|_{\text{steady}} \left\{ 1 - \exp\left(-(\omega_c t)^{3-\nu_A}\right) \right\} \tag{28}$$

where $v_f(X,t)\big|_{\text{steady}}$ represents Eq. (25). The corresponding general formula for the displacement is obtained by integrating Eq. (28) from 0 to $t$ and it can be written on the following form:

$$\frac{x_f(X,t)}{d} = \frac{E_A(X)}{(K_N + K_{gas})f_{\omega_c}(X,\xi)} \left\{ \frac{(\omega_c t)^{\nu_A - 1}}{\nu_A - 1} - \frac{1}{3 - \nu_A} \left[ \Gamma\left(\frac{\nu_A - 1}{3 - \nu_A}\right) - \Gamma\left(\frac{\nu_A - 1}{3 - \nu_A}, (\omega_c t)^{3-\nu_A}\right) \right] \right\} \tag{29}$$

where $\Gamma(a,z) = \int_z^\infty t^{a-1} \exp(-t)dt$ is the incomplete gamma function and $\Gamma(a,0) = \Gamma(a)$.

We now discuss the evolution of Eq. (28) and Eq. (29) as a function of the non-dimensional parameters $N_A$ and $T_A^* = T_A/T_{A0}$. First, it is useful to rewrite Eq. (28) and Eq. (29) by using these two parameters:

$$v_f^*(T_A^*, N_A) = \frac{v_f(T_A^*, N_A)}{l\omega_c} = T_A^* \left\{ 1 - \exp\left(-(N_A)^{3-\nu_A}\right) \right\}(N_A)^{\nu_A - 2} \tag{30}$$

and

$$x_f^*(T_A^*, N_A) = \frac{x_f(T_A^*, N_A)}{l} = T_A^* \left\{ \frac{(N_A)^{\nu_A - 1}}{\nu_A - 1} - \frac{1}{3 - \nu_A} \left[ \Gamma\left(\frac{\nu_A - 1}{3 - \nu_A}\right) - \Gamma\left(\frac{\nu_A - 1}{3 - \nu_A}, (N_A)^{3-\nu_A}\right) \right] \right\} \tag{31}$$

where $\nu_A = \nu_A(T_A^*)$. The two functions $v_f^*(T_A^*, N_A)$ and $x_f^*(T_A^*, N_A)$ represent the response of the system subject to the action $T_A^*$. According to Eqs. (30) and (31) nothing *a priori* suggests the response of these equations can be linear. This is a property commonly used in the analysis of experimental data. It is therefore interesting to observe the behavior of Eqs. (30) and (31) in this perspective. Fig. 1 shows first of all that the response of the present modeling is never linear at short times (i.e. for $t \le \omega_c^{-1}$) for both displacement and velocity. Moreover, it is also clear that during the transient regime the response function is quasi-independent of $T_A^*$. Fig. 1a shows that for $t > \omega_c^{-1}$ the linear approximation for the displacement is always possible when $T_A^* \le 1$ but this is no longer the case as soon as $T_A^*$ is greater than 1. Fig. 1b shows for $t > \omega_c^{-1}$ that the linear approximation for the velocity is only possible when $T_A^* \gg 1$. It can also be seen from Fig. 1b that the Newtonian regime is almost reached when $T_A^* \approx 100$. In the Newtonian regime it appears that the response function is also quasi-independent of $T_A^*$ at any time.

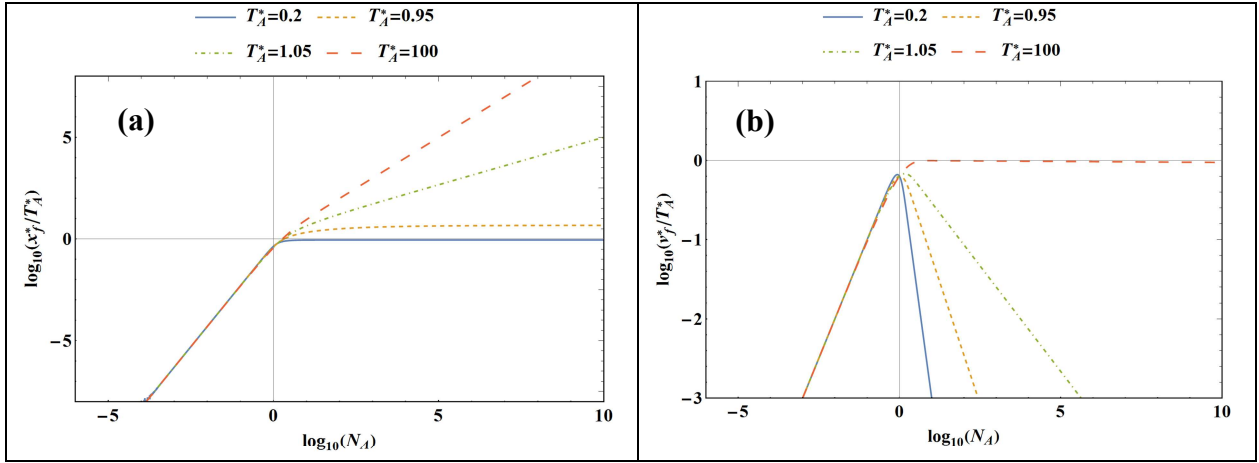

Fig. 1. Logarithmic plot versus the reduced duration for different values of the parameter $T_A^*$ of: (a) the displacement response Eq. (31); (b) the velocity response Eq. (30).

## 3 Application to liquid water viscoelastic data

The present modeling shows that the rheological behavior of the medium depends essentially on the value of $T_A^*$, i.e. of the ratio $E_A$ to $\left(K_N + K_{gas}\right) f_{\omega_c}\left(X, \xi\right)$. Table 2 shows that $E_A$ depends essentially on the stress or torque that is applied to the medium, i.e. on the configuration of the experimental set-up.

In the Newtonian limit, the denominator of $T_A^*$ depends only on $K_N$, $K_{gas}$, $\tau$ and $d$. In Ref. 11, it was shown that these quantities make it possible to define an expression of the dynamic viscosity which has allowed us to reproduce the experimental datasets on water in accordance with their uncertainties.

More generally, the quantity $\left(K_N + K_{gas}\right) f_{\omega_c}\left(X, \xi\right)$ contains the intrinsic properties of the medium as well as characteristic dimensions related to the experimental set-up. This quantity has the dimension of a stress that represents the threshold stress $\sigma_T$ (or the mechanical energy per unit volume threshold) for which $T_A^* = 1$, i.e. for which the system undergoes a "dynamic" phase transition according to Eq. (24). The determination of this threshold stress requires the knowledge of the function $f_{\omega_c}\left(X, \xi\right)$ which depends both on the properties of the medium and the experimental set-up.

Given that the various model parameters have already been determined in the case of water, the objective of this paragraph is then to determine an expression of $f_{\omega_c}\left(X, \xi\right)$ by analyzing liquid water relaxation data that show a threshold stress. Such water relaxation data were obtained using a plate-plate geometry with a gap $e = 125$ μm and a radius of the planar discs $R = 2$ cm. The details of the experimental set-up are published by Baroni *et al.* in Ref. 15 and the experimental data of Noirez *et al.* are published in Ref. 14.

The experimental conditions correspond to water at atmospheric pressure and a mean temperature of 294.65 K. For these conditions the 1995 IAPWS state equation formulation (Ref. 16) gives a liquid water density $\rho = 0.997886$ g/cm³. Considering $l = d = e$, then the fundamental parameters of the model can be calculated using the relations developed in Ref. 11. These parameters are grouped in Table 3.

| Name (unit) | Value |
|---|---|
| $K$ (GPa) | 2.95364 |
| $c_0$ (m/s) | 1720.43 |
| $\tau$ (ns) | 72.6558 |
| $N$ | $1.72788 \times 10^7$ |
| $v$ | 1.7174 |
| $H_N(v)$ | 217036 |
| $K_N$ (Pa) | 13609.2 |
| $K_{gas}$ (Pa) | 518.9 |
| $d_N$ (cm) | 2.002 |
| $\xi_0$ (Å) | 3.77906 |

Table 3. Numerical values of the fundamental parameters for liquid water at atmospheric pressure and 294.65 K when using the modeling from Ref. 11. The length $\xi_0$ represents the distance over which the fluctuations of the basic units are correlated in the bulk at rest and its expression is defined by Eq. (10) of Ref. 11.

Table 3 shows that the fluctuative distance $d_N$ corresponds exactly to the radius $R$ of the plates, taking $d = e$ as the dissipative distance. The two characteristic distances of the rheometer therefore appear "naturally" in the modeling, which demonstrates a high degree of consistency of the viscosity modeling presented in Ref. 11.

Also, in Ref. 11 it was shown that at atmospheric pressure and for a temperature between the triple point temperature and temperature corresponding to the saturated vapor pressure curve, the viscosity term due to the released gas is small in comparison with the liquid–like term equal to $K_N \tau$ : Table 4 shows that the viscosity corresponding to the released gas represents 3.8% of the total viscosity. Therefore, as a first approximation the term $K_{gas}$ can be neglected in comparison with $K_N$ in the next sections (a case where the term $K_{gas}$ is preponderant will be dealt in a future paper).

| Name (unit) | Value |
|---|---|
| $K_N \tau$ (mPa.s) from Table 3 | 0.98878 |
| $\eta$ (mPa.s) from Ref. 11 with $d = e = 125$ μm | 1.02649 |
| $\eta$ (mPa.s) from IAPWS08 (Ref. 17) | 0.96607 |
| $\eta_{Knu}$ (mPa.s) from Ref. 11 | 0.03770 |

Table 4. Different values for the dynamic viscosity of water at atmospheric pressure and 294.65 K. $\eta_{Knu}$ represents the viscosity of the released gas and its expression is defined by Eq. (17) of Ref. 11.

### 3.1. Evaluation of the threshold stress from relaxation experiment (at constant strain)

The relaxation data from Noirez *et al.* (Ref. 14) that will be analyzed hereafter were obtained by imposing a sudden displacement (modeled by a nearly Heaviside function of the strain which last about 0.03 s) of the moving planar disk measured at the distance $r = R$, which corresponds with the notations used at $x_f(R,e)/e = \gamma$. This relative displacement $\gamma$ (i.e. a shear strain amplitude) is given in percent. The torque is then measured to keep the imposed strain. Fig. 2 shows a typical example of the experimental results obtained upon applying a step strain to a 125 μm layer water. In spite of the important "noise" on the data, it can be observed first of all, a torque rise phase which corresponds to the shear stress during the initial displacement phase of the moving planar disk, then comes the stress decay phase, quite fast at first and then tends to stabilize to a constant value for enough longer times. This phase of decay corresponds to the relaxation phase itself. We will analyze the characteristics of these different phases more in detail later in this section. It can however be noticed that the times involved are very large compared to the characteristic macroscopic time $\tau$. The other important time in the system is $\omega_c^{-1}$ which will be determined below by the threshold stress analysis.

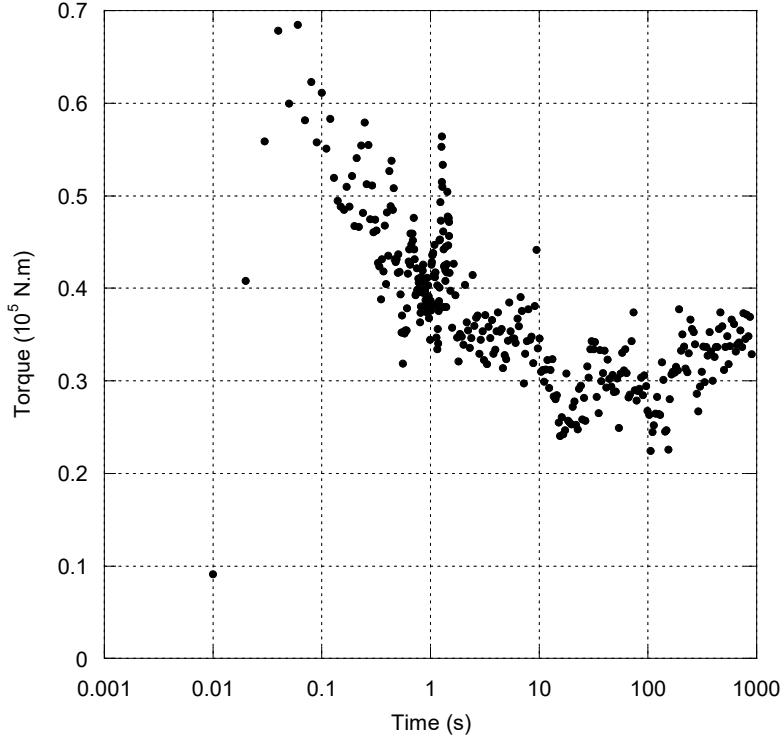

Fig. 2. Semi-logarithmic plot of the torque versus time measured for an imposed shear strain amplitude of 2.5 % on liquid water (shear stress relaxation experimental values recorded in plate-plate geometry – ARES2 measurements, gap thickness 125 μm, room temperature, total wetting conditions (alumina plates) from Ref. 14)

Now to compare these data with the threshold stress $\sigma_T$, the torque data are transformed (taking into account the device constants) into stress data by a law of proportionality. Since the stress in the medium does not depend on the altitude $z$ but only on the radial distance $r$ (disk-like geometry setup, see Ref. 4), the stress data obtained correspond to the stress at distance $r = R$ (i.e. to the maximum stress in the system) which is consistent with the measure of displacement (i.e. strain amplitude) imposed. An analysis of the threshold stress $\sigma_T$ by using these data in the form of a stress could then be done.

The longest time common in the experimental data being $t = 100$ s, we arbitrarily choose to analyze the shear stress versus shear strain for this duration. For this time value, it appears that most of the data shows a non-zero horizontal part of the relaxation dynamics or close to it, i.e. on the stress threshold value. Fig. 3 shows that this threshold stress (at $t = 100$ s) varies slightly as a function of the strain and has a bell shape with a maximum value for a shear strain around $\gamma = 100\%$. Given the parameters of the model, the only possible quantity compatible with these values is such that $f_{\omega_c}(X, \xi) = \xi_0/e$ where $\xi_0$ is the bulk correlation length defined in Table 3. It is remarkable to find that $\xi = \xi_0$ in the Newtonian regime.

Therefore, in the Newtonian regime, the threshold stress determined from relaxation data of liquid water can be expressed as follows:

$$\sigma_{T,\text{water}} = \left(K_N + K_{gas}\right)\frac{\xi_0}{e} \approx K_N\,\frac{\xi_0}{e} \qquad (32)$$

By taking the numerical values from Table 3, it is obtained the following shear stress value: $\sigma_{T,\text{water}} = 0.041\,\text{Pa}$.

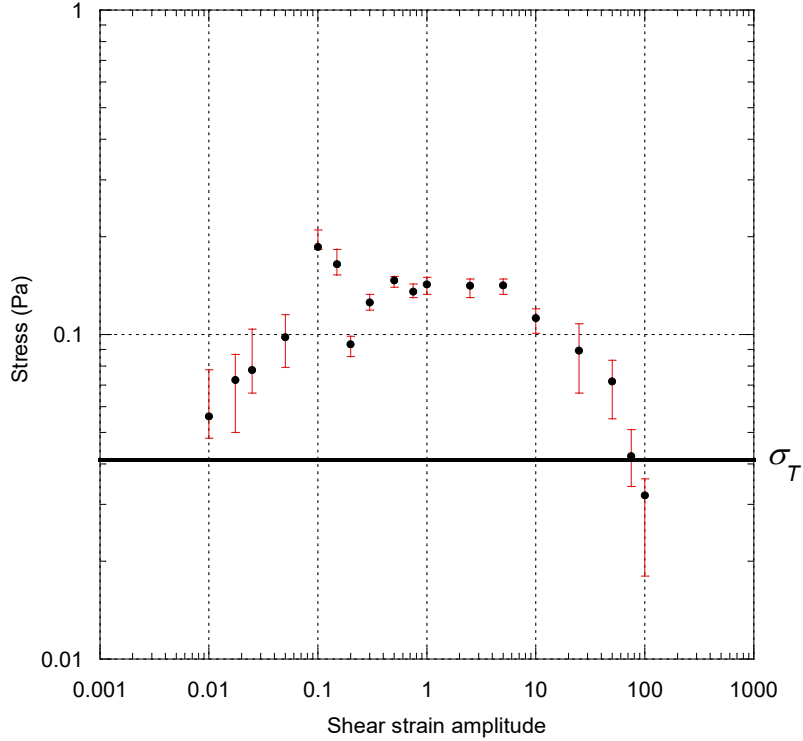

Fig. 3. Logarithmic plot of the measured stress versus the imposed shear strain amplitude for liquid water at a time $t = 100$ s (gap thickness 125 μm, room temperature measurements, alumina plate-plate geometry, data from Ref. 14).

It can be seen in Fig. 3 that $\sigma_T$ represents a low limit with respect to the experimental points. As it stands, the present modeling allows to define a threshold stress whose numerical value is compatible with the experimental results in a very large range of strain (i.e. 4 orders of magnitude). The threshold stress determined here represents a threshold stress of the bulk away from any influence of the walls.

We interpret the fact that in Fig. 3 the stress varies with the strain is partly related to the presence of the walls which modify the correlation length $\xi_0$ of the system according to the strain amplitude. This effect modifies the threshold stress as we will see in more detail in section 3.2.

The value of the correlation length $\xi_0$ (see Eq. (10b) of Ref. 11) is also consistent with *a priori* independent experimental results. Indeed, it has been shown for example in appendix C of Ref. 11 that the expression of $\xi_0$ fits perfectly with Xie *et al.*'s data (Ref. 18) obtained from the structure factor measured using synchrotron-based small-angle X-ray scattering in the supercooled phase of water.

The stress threshold allowed us to characterize the relaxation phase at long times. We now need to study the relaxation phase at short times as well as the stress rise phase. To do this, the following empirical expression will be used:

$$\sigma(t) = \underbrace{\exp\left(-\left(\frac{t_{\text{rise}}}{t}\right)^{3.237}\right)}_{\text{describes the rise phase}} \left\{ \underbrace{\sigma_0 \exp\left(-\frac{t}{\tau_\nu}\right)}_{\substack{\text{descibes the short times} \\ \text{relaxation phase}}} + \underbrace{\sigma_1 \left(1+\frac{t}{\beta\tau}\right)^{-\beta}}_{\substack{\text{describes the long times} \\ \text{relaxation phase}}} + \sigma_T \right\} \tag{33}$$

which describes well the stress evolution as a function of time for a given shear strain (see Fig. 4). In this equation the following parameters are fixed: $\sigma_T$, $\tau$ and $\tau_\nu$. The parameter $\tau_\nu$ represents the characteristic viscous diffusion time, i.e. $\tau_\nu = \dfrac{e^2}{\eta/\rho}$. Given the approximation $\eta \approx K_N \tau$, we can still write as a first approximation that $\tau_\nu \approx \tau\, H_N(\nu)$, i.e. all the characteristic times of the system are related to the fundamental characteristic time $\tau$. For the experimental conditions presented here the calculation gives $\tau_\nu = 16.139\,\text{ms}$. This time makes it possible to describe the relaxation phase at short times using a simple exponential law, i.e. the initial phase is governed by the viscous diffusion time to cross the sample. Eq. (33) shows however that a simple exponential decay does not allow to describe the whole relaxation kinetics but at long times, the data agree with a power law with an exponent globally much smaller than 1 (we can notice that this power law turns into an exponential only when the exponent tends towards infinity). This shows that this phase of decay is very slow and becomes slower as the strain amplitude is increased (i.e. it can be seen on the examples given in Fig. 4 that the values of $\sigma_1$ and $\beta$ decrease strongly as the strain amplitude increases).

The stress rise phase is governed solely by the rise time $t_{\text{rise}}$. It is then useful to compare these rise times with the characteristic times of the physical phenomena of the model. The analysis of the different data curves shows that the rise time varies little in such a way that, on average $t_{\text{rise}} \approx 7\,\text{ms}$. But since the value of the maximum stress $\sigma_{\max}$ globally increases as the strain increases, we deduce that the total rise time to reach this maximum stress increases as the strain increases, which seem physically understandable. It appears that the time taken by the experimental device to achieve the imposed strain is at least of the same order of magnitude as $\tau_\nu$ and is very long in front of $\tau$.

Moreover, we can also compare $t_{\text{rise}}$ to the transient time $\omega_c^{-1} = 24.032\,\text{ms}$. In other words, by the time the system was allowed to relax, the shear information had time to pass through the entire sample; also, the viscous diffusion had time to diffuse through the thickness of the sample. It is also noted that the beginning of the relaxation phase coincides with the end of the transient regime so that Eq. (23) is appropriate to analyze this phase.

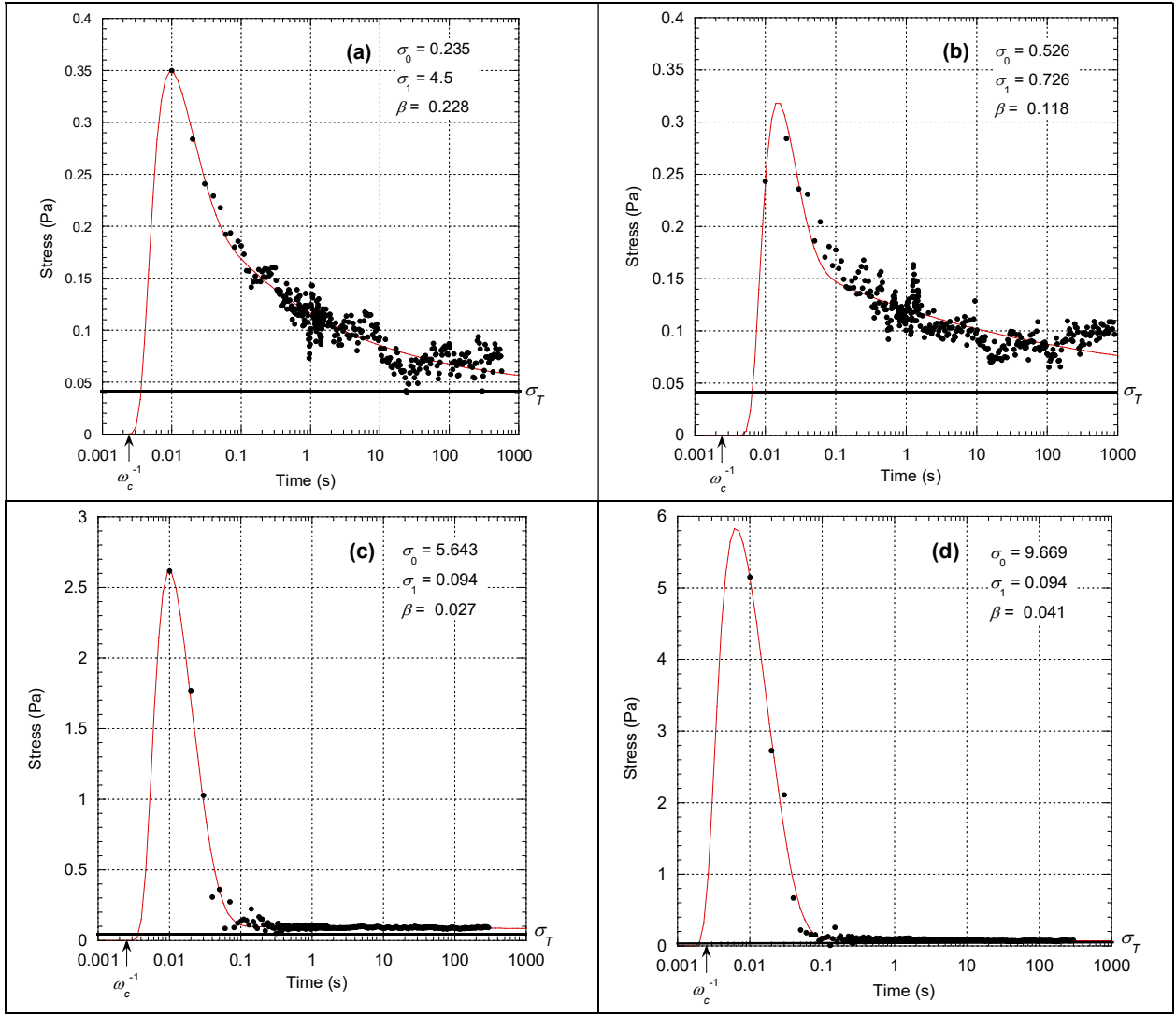

Fig. 4. Semi-logarithmic plot of some stress data from Ref. 14 versus time (black points) with Eq. (32) (red curve) for different values of the imposed shear strain amplitude of liquid water: (a) 1.75%, (b) 2.5%, (c) 2500% and (d) 5000%. $\sigma_0$ and $\sigma_1$ are given in Pa. Gap thickness 125 μm, room temperature measurements, alumina plate-plate geometry.

## 3.2. Determination of the shear elastic modulus from relaxation data

The relaxation data show that the medium tries to return to equilibrium following the action performed. In other words, the basic units try to return to an equilibrium position when the displacement of the moving plate is stopped. Determining this displacement, knowing the applied stress, makes it possible to define a shear elastic modulus $G$ of the medium such that:

$$G = \sigma/\gamma \quad \text{with} \quad \gamma = x_f(R,e,\sigma)/e \tag{34}$$

where $x_f(R,e,\sigma)$ determines the displacement corresponding to the applied stress $\sigma$. Indeed, according to Table 2(d), it appears that $E_A(R,e) = \dfrac{2\Gamma}{\pi R^3}$ but this expression is also equal to the measured stress $\sigma$ on the moving planar disk, therefore $E_A$ is determined directly from the experimental data.

We now focus on the relaxation phase at long times, i.e. the region for which we can consider that the stress almost does not vary anymore. The stress corresponding to this region will be noted $\sigma_\infty$.

To describe the system in this region, one can then use Eq. (29) but to perfectly account for the relaxation data, a parameter $\xi^*$ must be introduced such that $f_{\omega_c}(e, \xi) = \xi^* \xi_0 / e$ and $\xi^* \xi_0$ represents an effective coherence length $\xi$ that reflects the effect of walls in the system and also possible "defects". $\xi^*$ is the only free parameter of the model. The non-dimensional parameter $\xi^*$ is determined so as to achieve the largest shear strain amplitude in the range $[0, \gamma_0]$ where $\gamma_0$ represents the imposed shear strain amplitude. Indeed, by taking a simple image of Maxwell's model, the total displacement of a fluid particle is the result of both elastic deformation and viscous flow. Whatever the strain amplitude, energy dissipation takes place but for small enough deformations, one generally neglects the viscous dissipation in front of the elastic deformation. This is indeed what is obtained for water as long as $\gamma_0$ is less than or equal to 100%. Beyond this strain amplitude, it appears for water that the viscous displacement can no longer be neglected and becomes more pronounced as the shear strain amplitude increases. For $\gamma_0 > 100\%$, it is no longer possible to find values of the parameter $\xi^*$ such that Eq. (29) satisfies both the imposed strain amplitude $\gamma_0$ and the stress value $\sigma_\infty$. It is also not possible to consider strain amplitude values much greater than 100%. Given that $\sigma_\infty$ is maximum (see Fig. 3) and $T_A^* \approx 1$ (see Fig. 6b) for $\gamma_0 = 100\%$, it appears "natural" to consider that for imposed shear strain amplitude $\gamma_0 > 100\%$, the maximum shear strain corresponding to $\sigma_\infty$ is such that $\gamma = 100\%$. In other words, for this water relaxation experiment, $G$ is determined as:

$$\begin{cases} G = \sigma_\infty / \gamma_0, \text{ if } \gamma_0 \leq 100\% \\ G = \sigma_\infty / \gamma \text{ with } \gamma = 100\%, \text{ otherwise} \end{cases} \quad (35)$$

Fig. 5 shows that the application of Eq. (35) to the relaxation water data makes it possible to define a constant shear elastic modulus $G$ at long times over at least 1 to 2 decades.

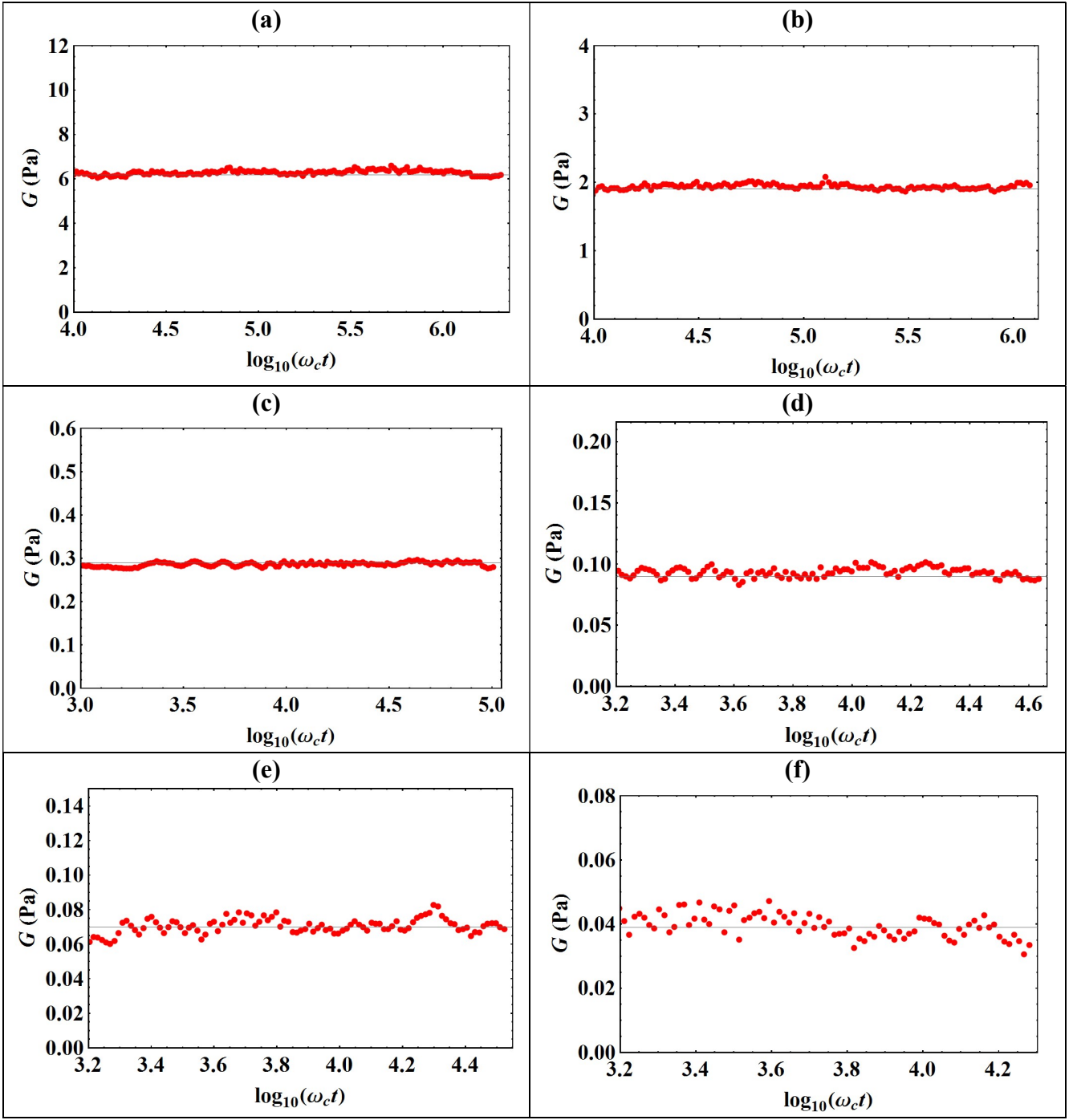

Fig. 5. Application examples of Eq. (34) to liquid water relaxation data from Ref. 14 for different values of the imposed shear strain amplitude: (a) 1%, (b) 5%, (c) 50%, (d) 2500%, (e) 5000% and (f) 7500%. Gap thickness 125 μm, room temperature measurements, alumina plate-plate geometry.

Fig. 6a shows that the value of the shear modulus of elasticity $G$ decreases continuously with increasing shear strain amplitude, which implies a very low value of this modulus when the Newtonian regime is reached according to the definition of this regime. On this figure, it is also shown what can be deduced from the solution of Maxwell viscoelastic model for an imposed Heaviside strain (i.e. $\dfrac{d\sigma}{dt} + \dfrac{G_{\text{Maxwell}}}{\eta}\sigma = 2G_{\text{Maxwell}}\gamma_0\delta(t)$ where $\delta(t)$ is the Dirac delta function), for which $G_{\mathrm{Maxwell}}$ is given by the relation $G_{\mathrm{Maxwell}} = \dfrac{\sigma_{\max}}{2\gamma_0}$ where $\sigma_{\max}$ represents the value of the stress peak; this maximum value is determined here from Eq. (33). It can be observed that the order of magnitude of the numerical values as well as the variations are quite similar for $\gamma_0 \leq 100\%$. On the other hand, beyond 100% the Maxwell's model seems to lead to a constant value $G_{\mathrm{Maxwell}} \approx 52.92\,\mathrm{mPa}$ which is close to the threshold stress value given by Eq. (32). But in addition, the characteristic relaxation times of the Maxwell's model which correspond to an exponential decrease of the stress such that $t_{\mathrm{Maxwell}} = \eta/G_{\mathrm{Maxwell}}$ are incompatible with the data for shear strain amplitude $\gamma_0$ below 100%. Indeed, for small shear strain amplitudes, $t_{\mathrm{Maxwell}}$ is much smaller than one millisecond whereas we have previously shown that the exponential decay is governed by the time $\tau_\nu$ which is of the order of 16 ms. Maxwell's model is therefore too simple or simply not suitable for analyzing these relaxation data even though it appears that the values of $G_{\mathrm{Maxwell}}$ are physically acceptable.

From Eq. (35), it is then deduced that $\xi^*$ is solution of the following relation:

$$\begin{cases} \gamma_0 = x_f\left(R, e, \sigma_\infty, \xi^*\right)/e, \text{ if } \gamma_0 \leq 100\% \\ \gamma = x_f\left(R, e, \sigma_\infty, \xi^*\right)/e, \text{ with } \gamma = 100\%, \text{ otherwise} \end{cases} \tag{36}$$

where the expression of $x_f$ is given by Eq. (29). Fig. 6b shows that the correlation length keeps decreasing until $\xi$ reaches the value $\xi_0$ for $\gamma_0 = 10\,000\%$. It therefore appears logically that for large shear strain such that $T_{\mathrm{A}}^*$ is close to the Newtonian regime (i.e. $T_{\mathrm{A}}^* > 100$), we find $\xi \approx \xi_0$ and therefore the threshold stress $\sigma_T$ of the Newtonian regime is given by Eq. (32). It is also found that the transition from the solid-like to the liquid-like regime occurs for $\gamma_0 = 100\%$, which is consistent with the basis of Eq. (35).

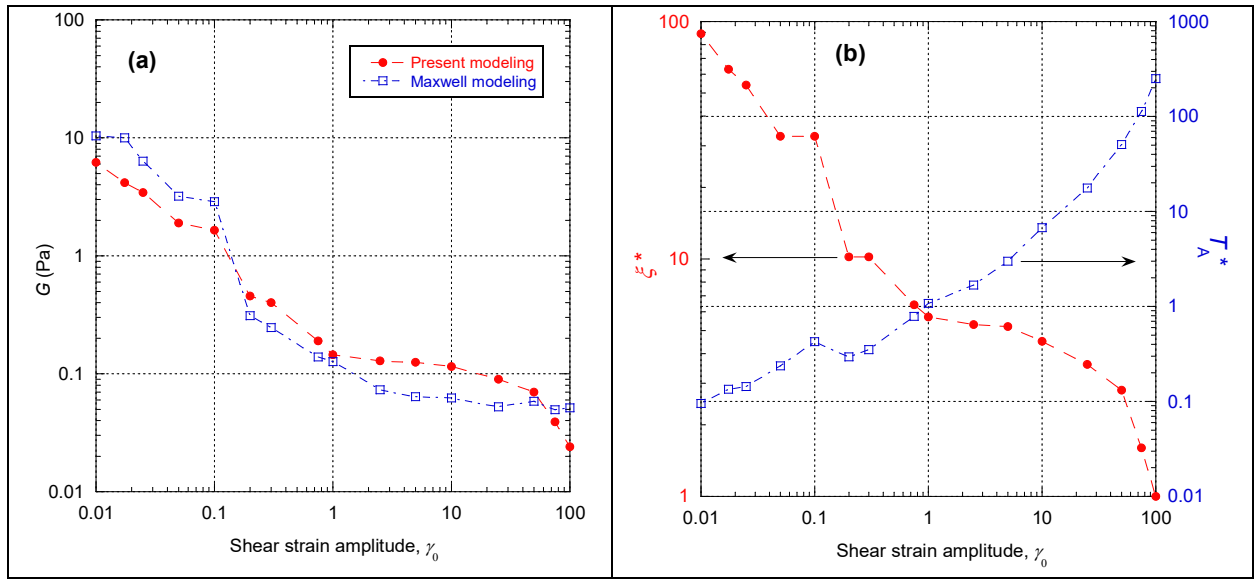

Fig. 6. Logarithmic plots as function of the imposed shear strain amplitude $\gamma_0$: (a) shear elastic modulus comparison of the present modeling with the Maxwell's solution for an imposed Heaviside strain; (b) variations of the non-dimensional correlation length $\xi^* = \xi/\xi_0$ (left axis) and the non-dimensional action temperature $T_{\mathrm{A}}^* = \sigma_{\max}\Big/\left(K_N\,\dfrac{\xi}{e}\right)$ (right axis). The dashed lines are eyes guides.

The extrapolation of the viscoelastic properties of water to small and large strains depends on the extrapolation that can be imagined for $\xi(\gamma_0)$. Considering that the stress $\sigma_\infty$ corresponds to long times such that $t \to \infty$ and knowing that $v_A\left(\dfrac{\sigma_\infty}{K_N \xi/e}\right)$ is a negative function, then Eq. (29) can be written in this limit more simply as:

$$\left(\frac{x_f(X,t)}{e}\right)_{t\to\infty} = \frac{\sigma_\infty}{\left(K_N + K_{gas}\right)\dfrac{\xi}{e}}\left\{-\frac{1}{3-v_A}\Gamma\left(\frac{v_A-1}{3-v_A}\right)\right\} \tag{37}$$

By defining $\sigma_\infty^* = \dfrac{\sigma_\infty}{\sigma_T}$, Eq. (37) can simply be written:

$$\gamma_{t\to\infty} = \frac{\sigma_\infty^*}{\xi^*}\frac{1}{v_A\left(\sigma_\infty^*/\xi^*\right)-3}\Gamma\left(\frac{v_A\left(\sigma_\infty^*/\xi^*\right)-1}{3-v_A\left(\sigma_\infty^*/\xi^*\right)}\right) \tag{37bis}$$

If we assume for large strains that $\xi^* = 1$ and that $\gamma_{t\to\infty}$ remains locked at 100%, then the inversion of Eq. (37bis) leads us to a limit value for $\sigma_\infty^*$ such that $\sigma_\infty^* = 0.618$, i.e. $\sigma_\infty = 25.42\,\text{mPa}$. It can be seen that this value is perfectly compatible with the stress value corresponding to $\gamma_0 = 10\,000\%$ in Fig. 3. According to Eq. (35), this $\sigma_\infty$ value is also the limit value of $G$. This value of $G$ is also compatible with the value obtained for $\gamma_0 = 10\,000\%$ in Fig. 6a. For this limit, $K_N \xi/e \to \sigma_T$ and thus $\sigma_T$ represents the threshold stress in the Newtonian regime. In other words, there is no such thing as a "perfect" fluid, i.e. one that flows under the action of an arbitrarily weak external stress.

Now for small strains, we have according to Eq. (36) $\gamma_{t\to\infty} = \gamma_0$ and according to Fig. 6b we can admit that the limit when $\gamma_0 \to 0$ is such that $\xi \to e$ (i.e. $\xi^* \to e/\xi_0 = 330\,770$). For this limit, Eq. (37bis) implies that $\sigma_\infty^* \to 0$. This limit is consistent with the observed stress variations in Fig. 3. Moreover, this limit seems to be physically acceptable since if there is no strain, there is also no relaxation and therefore no long-time stress. According to Eq. (35), it can be deduced that $G$ has a finite value when $\gamma_0 \to 0$ but this value cannot be determined here. For this limit, $K_N \xi/e \to K_N$ and thus $K_N$ represents the threshold stress in the zero-strain limit.

### 3.3. Determination of the shear elastic and viscous moduli from small-amplitude oscillatory motion

The same experimental device as for the stress relaxation study was used by imposing a low-amplitude oscillatory motion on the moving planar disk to study the frequency response of the medium (Refs. 4 and 14). We use these Noirez *et al.*'s data to carry out a linear analysis of the low amplitude of oscillations. In the linear analysis framework, the function $G(t)$ is interpreted as the linearized impulse response of the shear medium. The Fourier-Laplace transform of the impulse response $G(t)$ leads to the expression of the conventional terms of shear elastic ($G'$) and viscous ($G''$) moduli such that:

$$G'(\omega) = \text{Im}\left[ \omega \int_0^\infty G(t) e^{i\omega t} dt \right] \quad \text{and} \quad G''(\omega) = \text{Re}\left[ \omega \int_0^\infty G(t) e^{i\omega t} dt \right] \tag{38}$$

where $\text{Im}[\bullet]$ and $\text{Re}[\bullet]$ represent the imaginary and real part respectively. Since $G(t)$ is a real function such that $G(t) = 0$ for $t < 0$, then $G'(\omega)$ and $G''(\omega)$ satisfy Kramers-Kronig (K-K) relations.

As long as the resulting torque function keeps the shape of the imposed strain wave, then the above formalism applies and the experimental set-up allows to directly extract the functions $G'(\omega)$ and $G''(\omega)$. These experimental results have already been published in Ref. 14.

In this section, we will therefore develop a model for the relaxation function $G(t)$ which satisfies the constraints of the linear approach and which relies on all the analysis developed previously to study the frequency responses of the functions $G'(\omega)$ and $G''(\omega)$ for water.

For the analysis of the relaxation data, we have seen that two characteristic times $\tau$ and $\tau_\nu$ are involved in an exponential law and in a power law which is nothing but a generalized exponential since, when the exponent tends towards infinity, a simple exponential law is recovered. The exponential law as well as the generalized exponential having the correct properties at $t = 0$, we will therefore keep these functions to describe the relaxation function $G(t)$.

In addition to these two functions that describe relaxation at long and very long times, what happens at short times must also be specified. At short times, it is necessary that the relaxation information crosses through the basic units which have the characteristic length scale $q_c^{-1}$. Therefore, we will describe the short time relaxation function $G(t)$ by an exponential law with the characteristic time $\tau_{\text{mi}} = \dfrac{1}{c_0 q_c} = \dfrac{\tau}{d\, q_c}$.

Finally, we assume that the relaxation function $G(t)$ can be written as follows:

$$G(t) = \underbrace{(K - K_N)\exp\left(-\frac{t}{\tau_{mi}}\right)}_{\text{short time relaxation}} + \underbrace{(K_N - \sigma_\xi)\exp\left(-\frac{t}{\tau}\right)}_{\text{long time relaxation}} + \underbrace{\sigma_\xi\left[1 + \frac{t}{\alpha\tau_\nu}\right]^{-\alpha}}_{\text{very long time relaxation}} \tag{39}$$

where $\sigma_\xi = K_N \dfrac{\xi}{e} = \sigma_T\, \xi^*$. It is assumed at $t = 0$ that $G(0) = K$, which is consistent with $\tau_{\text{mi}}$ since $K$ represents the shear elasticity between the basic units. Then at longer times the whole system is affected and therefore $K_N$ represents the global shear elasticity and then the threshold is reached at very long times. The exponent $\alpha$ and the correlation length $\xi$ are determined from the $G'(\omega)$ experimental dataset.

The frequency response is obtained by numerically integrating Eq. (39) into Eq. (38). The smallest shear strain amplitude studied for water is here $\gamma_0 = 2.5\%$ at a mean temperature of 294.65 K (i.e. same average temperature as for the corresponding relaxation data). Fig. 7 shows that the variation of $G'$ as well as its amplitude can be perfectly reproduced with values of the parameters $\alpha$ and $\xi^*$ consistent with those determined for the relaxation data. The variation of $G'$ with $\omega$ corresponds here to a very weak slope which necessarily implies that $G''$ have quasi-zero values if they respect the K-K relations. However, $G''$ experimental data in Fig. 7 seem to indicate that they violate these relations. In fact, the discrepancy between the experimental data and the theoretical curve is simply due to the fluctuations of $G$ which do not have to verify the K-K relations since they exist at all times and these fluctuations contribute to the dissipation when there is a flow due to an out-of-equilibrium situation.

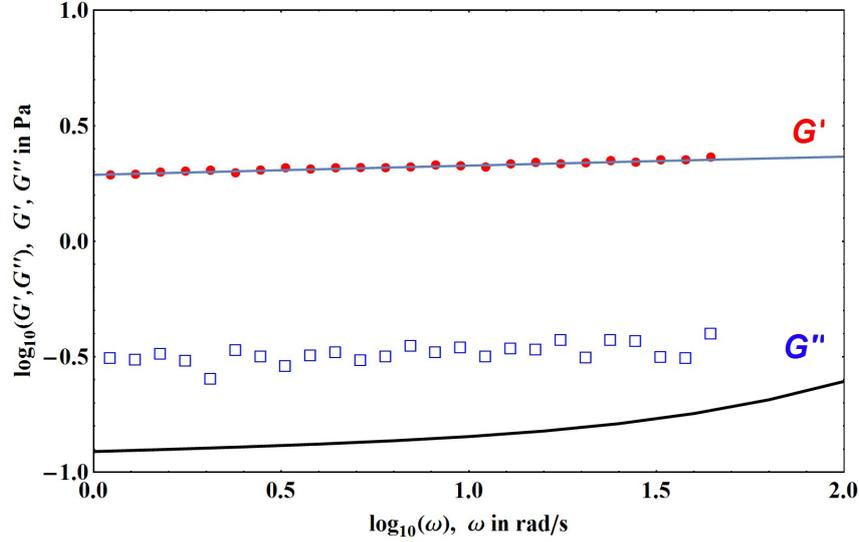

Fig. 7. Logarithmic plot of $G'(\omega)$ and $G''(\omega)$ experimental data (red points and open blue squares from Ref. 14) with the present modeling (blue and black curves) for $\gamma_0 = 2.5\%$ and a mean temperature of 294.65 K. $\alpha = 0.04$ and $\xi^* = 62$. Gap thickness 125 μm, alumina plate-plate geometry.

To estimate the excess of $G''$ due to these fluctuations, it is necessary to estimate their Root Mean Square amplitude (RMS). An estimate of the latter can be obtained by subtracting Eq. (33) from the experimental relaxation data and then the RMS value of this noise is calculated. For $\gamma_0 = 2.5\%$, the mean value of the RMS noise value is 0.331 Pa. Fig. 8 shows that the present modeling can perfectly reproduce the evolution of $G''(\omega)$ by adding a constant value $\Delta G'' = 0.185$ Pa to this function. This value is lower than the one determined with the relaxation data, which is not abnormal inasmuch as the experiments are different but the order of magnitude is quite coherent.

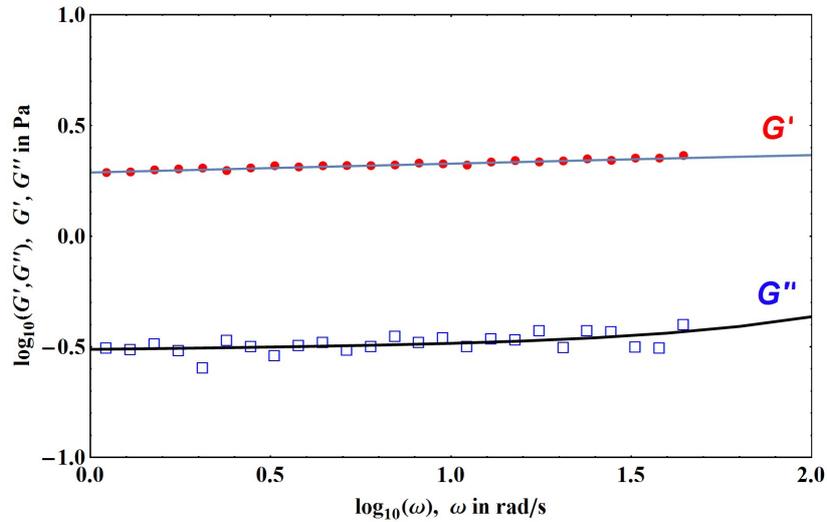

Fig. 8. Logarithmic plot of $G'(\omega)$ and $G''(\omega)$ experimental data (red points and open blue squares from Ref. 14) with the present modeling for which an excess of $G''$ has been added (blue and black curves). $\gamma_0 = 2.5\%$, the mean temperature is 294.65 K, $\alpha = 0.04$, $\xi^* = 62$ and $\Delta G'' = 0.185$ Pa. Gap thickness 125 µm, alumina plate-plate geometry.

Fig. 9 shows the extrapolation of the present model at high frequencies: it can be observed that $G'$ has three levels corresponding to the three distinct elastic constants of the model. We will see later that these levels are consistent with other kinds of experimental data.

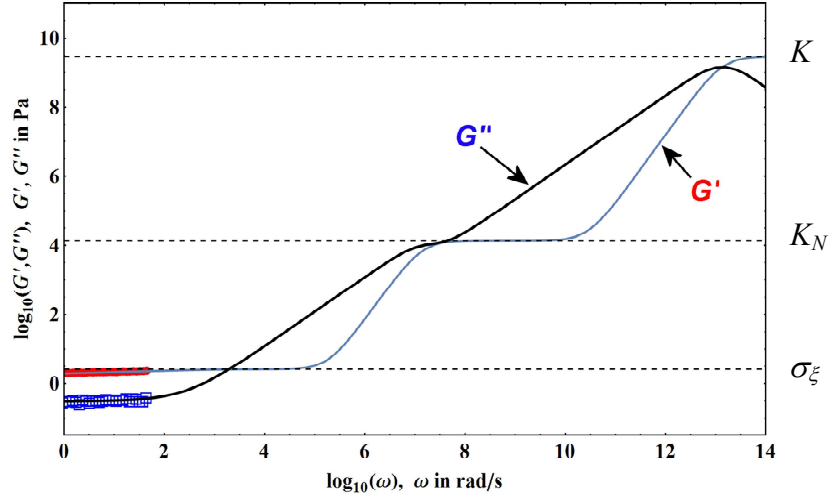

Fig. 9. Logarithmic plot of $G'(\omega)$ and $G''(\omega)$ experimental data (red points and open blue squares from Ref. 14) with the present modeling extrapolated to high frequencies (blue and black curves) for $\gamma_0 = 2.5\%$ and a mean temperature of 294.65 K. $\alpha = 0.04$, $\xi^* = 62$ and $\Delta G'' = 0.185$ Pa. Gap thickness 125 µm, alumina plate-plate geometry.

We will analyze in the same way the other data from Noirez *et al.* of $G'(\omega)$ and $G''(\omega)$ with increasing shear strains. The next strain amplitude which has been studied corresponds to $\gamma_0 = 5\%$ for a mean temperature of 294.85 K. Fig. 10 shows that by taking into account a constant dissipation excess $\Delta G'' = 0.155$ Pa, the present modeling allows to reproduce correctly the experimental data except at low frequencies. The fact that $G''$ increases while $G'$ decreases when the frequency is decreased indicates that this effect is due to noise which increases strongly at low frequencies and can no longer be *a priori* assimilated to a constant value. For the relaxation data corresponding to $\gamma_0 = 5\%$, the mean value of the RMS noise values is 0.146 Pa which is perfectly consistent with the value of $\Delta G''$.

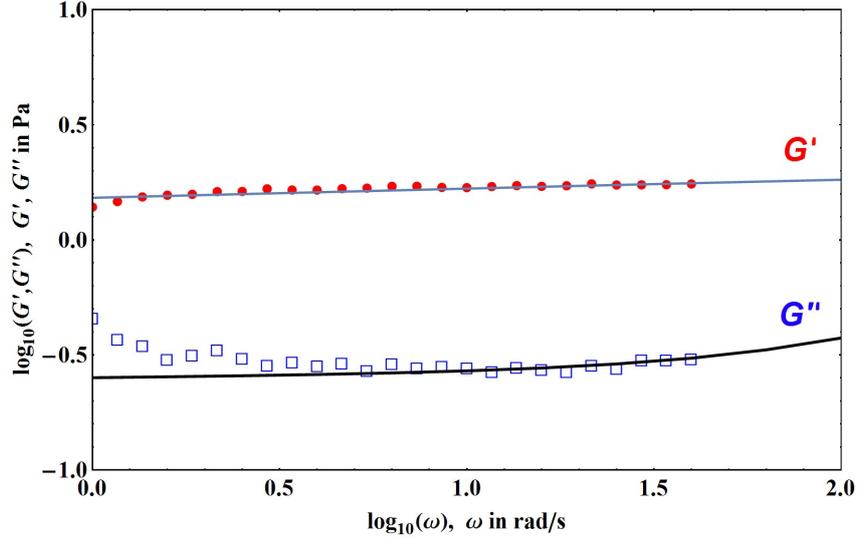

Fig. 10. Logarithmic plot of $G'(\omega)$ and $G''(\omega)$ experimental data (red points and open blue squares from Ref. 14) with the present modeling for which an excess of $G''$ has been added (blue and black curves). $\gamma_0 = 5\%$, the mean temperature is 294.85 K, $\alpha = 0.04$, $\xi^* = 49$ and $\Delta G'' = 0.155$ Pa. Gap thickness 125 µm, alumina plate-plate geometry.

The next strain amplitude which has been studied corresponds to $\gamma_0 = 30\%$ for a mean temperature of 294.33 K. Fig. 11 shows that by taking into account a constant dissipation excess $\Delta G'' = 0.065$ Pa, the present modeling allows to reproduce correctly the experimental data. For the relaxation data corresponding to $\gamma_0 = 30\%$, the mean of the RMS noise values is 0.013 Pa which is again consistent with the value of $\Delta G''$. Here the torque signal is slightly distorted compared to the strain signal and therefore linear analysis is not strictly allowed. However, this effect is mainly reflected in the numerical value of the parameters.

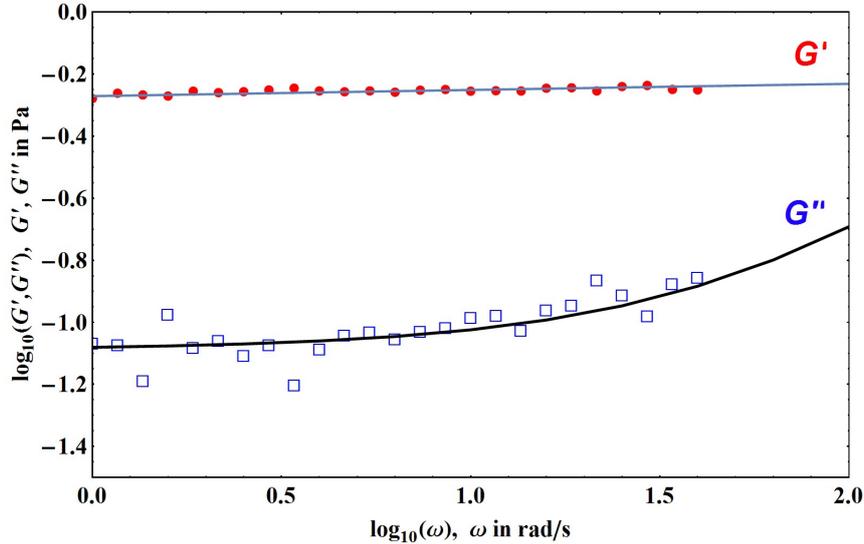

Fig. 11. Logarithmic plot of $G'(\omega)$ and $G''(\omega)$ experimental data (red points and open blue squares from Ref. 14) with the present modeling for which an excess of $G''$ has been added (blue and black curves). $\gamma_0 = 30\%$, the mean temperature is 294.33 K, $\alpha = 0.02$, $\xi^* = 15$ and $\Delta G'' = 0.065$ Pa. Gap thickness 125 µm, alumina plate-plate geometry.

For strain amplitudes greater than 30%, the torque signals are too distorted compared to the sinusoidal strain signal to be analyzed linearly. The torque recovers a sinusoidal shape for $\gamma_0 = $ 7500%. Fig. 12 shows first of all that here $G''$ has become greater than $G'$ while $G'$ is practically horizontal (i.e. $\alpha = 0.01$). In other words, the values of $G''$ are entirely determined by the fluctuations such as $\Delta G'' = 0.0022$ Pa. For the relaxation data corresponding to $\gamma_0 = 7500\%$, the mean of the RMS noise values is 0.00013 Pa. The value of $\Delta G''$ is relatively high compared to the relaxation data but this is consistent with the other amplitudes which show that the fluctuations in the oscillating experiments are globally greater than in the relaxation ones. For this frequency range, $G''$ is not linear with the frequency but has a slight curvature.

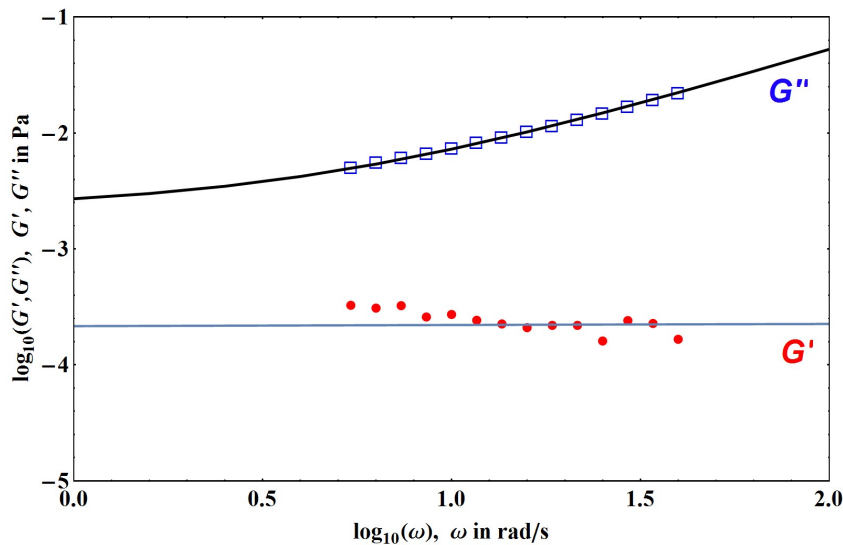

Fig. 12. Logarithmic plot of $G'(\omega)$ and $G''(\omega)$ experimental data (red points and open blue squares from Ref. 14) with the present modeling for which an excess of $G''$ has been added (blue and black curves). $\gamma_0 = 7500\%$, the mean temperature is 294.96 K, $\alpha = 0.01$, $\xi^* = 1$ and $\Delta G'' = 0.0022$ Pa. $\tau_{mi}$ is here multiplied by a factor of 1.3 while $K$ is multiplied by a factor of 0.34. Gap thickness 125 μm, alumina plate-plate geometry.

For the 7500% strain amplitude, the agreement between the present modeling and the data is obtained by "artificially" strongly decreasing the value of $K$ and slightly increasing the value of $\tau_{mi}$ by a multiplicative factor. This indicates that the system has undergone major modifications and in particular that there has probably been slippage on the walls. Fig. 13 shows what is obtained if the multiplicative factors of $\tau_{mi}$ and $K$ are reset to 1: it can be observed that the values of $G''$ are little modified in the frequency range corresponding to the data while the value of $G'$ is strongly increased in the same range; the value obtained for $G'$ is then practically identical to the value of $G$ determined from relaxation data (see Table 5) which shows the coherence of the analysis.

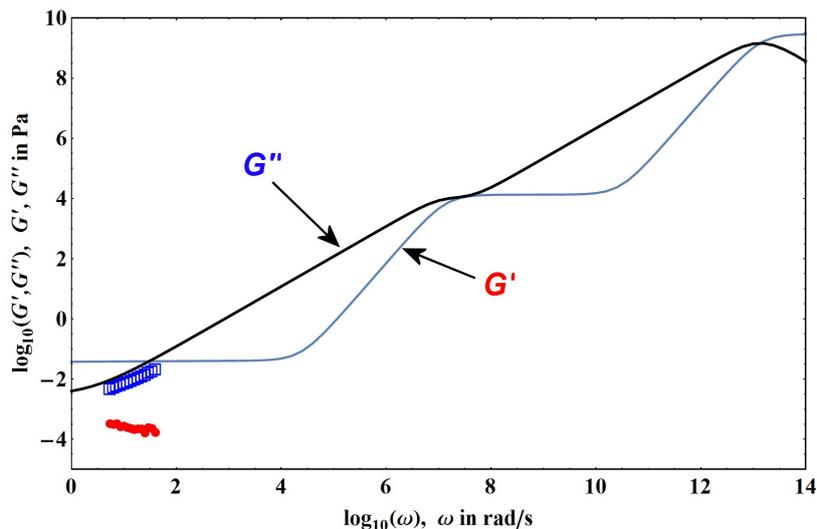

Fig. 13. Logarithmic plot of $G'(\omega)$ and $G''(\omega)$ experimental data (red points and open blue squares from Ref. 14) with the present modeling for which an excess of $G''$ has been added (blue and black curves). $\gamma_0 = 7500\%$, the mean temperature is 294.96 K, $\alpha = 0.01$, $\xi^* = 1$ and $\Delta G'' = 0.0022$ Pa. Gap thickness 125 µm, alumina plate-plate geometry.

It is useful at this stage to make a comparison between some significant results obtained with the two types of analysis. Fig. 14 shows the evolution of the $\xi^*$ variable as a function of the strain amplitude for the two kinds of experiments: it can be observed that the two theoretical analysis presented give perfectly coherent results for the evolution of the correlation length in the system.

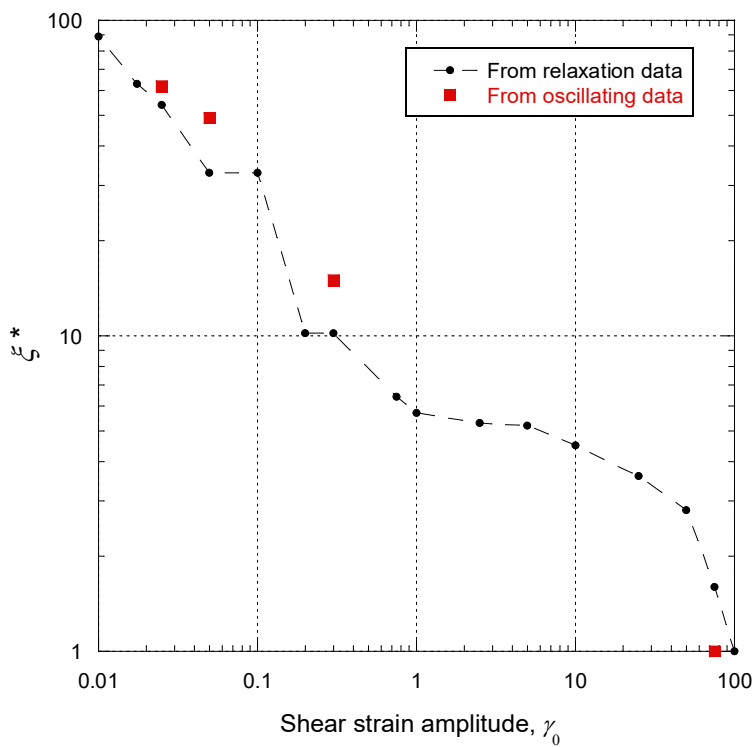

Fig. 14. Logarithmic plots of the non-dimensional correlation length $\xi^* = \xi/\xi_0$ as function of the imposed shear strain amplitude $\gamma_0$ for the two types of analysis corresponding to two experimental configurations. The dashed line is eyes guide.

Taking into account the fact that the relaxation and oscillating experiments are quite different with respect to the perturbation of the medium, Table 5 shows that the values deduced by the two theoretical analysis of $G$ and $G'$ are very consistent.

| $\gamma_0$ (%) | $G$ from relaxation data (Pa) | $G'$ from oscillating data (Pa) |
|---|---|---|
| 2.5 | 3.45 | 1.939 |
| 5 | 1.9 | 1.524 |
| 30 | 0.4 | 0.536 |
| 7500 | 0.039 | 0.0375 |

Table 5. Comparison of the numerical values of $G$ and $G'$ for water at atmospheric pressure obtained at different imposed shear strain amplitude $\gamma_0$ for the two types of analysis corresponding to two experimental configurations. The values of $G'$ are calculated for the smallest frequency (i.e. 1 rad/s) in the range of the experimental data.

It is now interesting to complete the above analysis for much higher frequencies. The data from Bund *et al*. (Ref. 19) concern water-glycerol mixtures at different concentrations and allow extrapolation of what should be obtained for pure water at the single frequency of 10.042 MHz. The interest of the Bund *et al*.'s data comes from the fact that the volume of liquid used is very similar to the volume of water in the plate-plate rheometer of the experiments described previously concerning Noirez *et al*.'s data: 0.2 cm$^3$ for Bund *et al.*'s experiment and 0.157 cm$^3$ for the plate-plate rheometer. On the other hand, the geometry is very different, as is the experimental set-up: the resonator radius is $R = 0.75$ cm so we can deduce an equivalent uniform cylindrical thickness $e = 1131.77$ μm of the sample. Therefore, the sample thickness here is ten times greater than in the plate-plate rheometer. For the corresponding frequency and a temperature of 293.65 K, the linearly extrapolated value of $G'$ for water is $G' = 18\,000$ Pa with an uncertainty of 12%. Considering the results of the two types of resonators used, a linear extrapolation of the tangent of mechanical loss angle $\tan(\delta) = G''/G'$ gives values between 0.639871 and 0.635414. By taking into account the value of $G'$, this leads to consider that the value of $G''$ must be between 11 517.7 Pa and 11 437.5 Pa. This uncertainty can be greatly expanded if the uncertainty on $G'$ is taken into account.

The parameters of the model are essentially determined by the geometry of the experiment. In particular, it has been previously shown that the fluctuative distance $d_N$ corresponds approximately to the radius $R$ and the dissipative distance $d$ corresponds to the thickness $e$. Due to the high frequency, the value of $\alpha$ can be arbitrarily set. Similarly, the value of $\xi^*$ mostly affects low frequencies and therefore has practically no influence for the frequency studied here. A value $\xi^* = 1$ can be set since the strain amplitude is not known for this experiment.

Fig. 15 shows that the value of $G'$ is very close to the expected value (i.e. 17 205.7 Pa) and therefore well within the uncertainty. On the other hand, we note that the value of $G''$ is higher than the expected value and outside the uncertainty. The discrepancy may be due to different effects such as linear extrapolation which may be incorrect. The value may also be higher than expected due to the presence of fluctuations as in the rheometer experiment and which must exist and which contribute to a strong increase in dissipation.

The only way to obtain a value of $G''$ in agreement with the expected value is to slightly reduce the value of $G(t=0)$ by a factor of 0.83. The surface of the resonators was covered with gold, while in the case of the plate-plate rheometer, the plates are made of alumina. Therefore, the wetting of the surface by water is very different in the two experiments and a lower wetting in the case of the Bund *et al.*'s experiment goes in the direction of a decrease of $G(t=0)$.

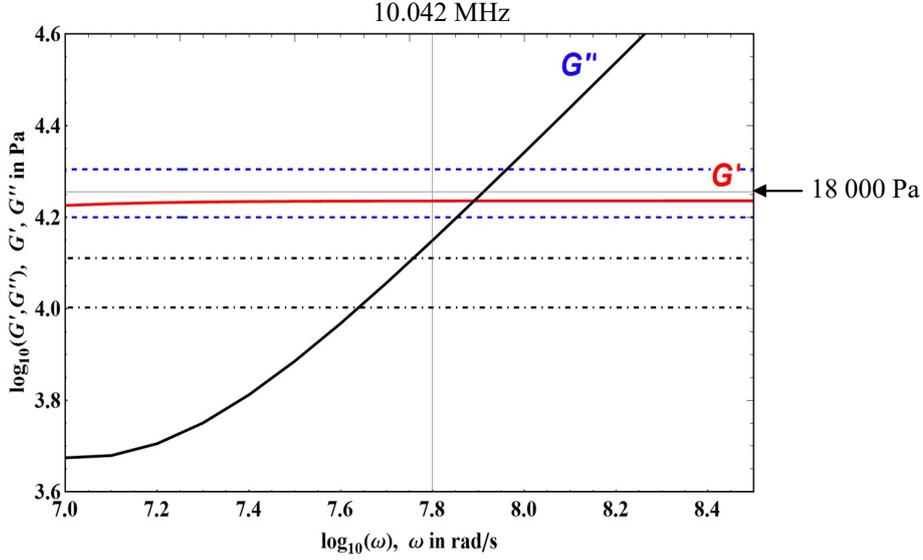

Fig. 15. Logarithmic plot of $G'(\omega)$ (red curve) and $G''(\omega)$ (black curve) calculated with the present modeling in a small frequency range. The blue dashed lines represent an uncertainty of 12% around the expected value of $G'$ and the black dot-dashed lines represent an uncertainty of 12% around the expected value of $G''$. $T = 293.65$ K, $\alpha = 0.04$, $\xi^* = 1$.

Fig. 16 shows that the value of $G'$ for Bund *et al.*'s experiment corresponds to the level at $K_N$ which justifies the existence of this level.

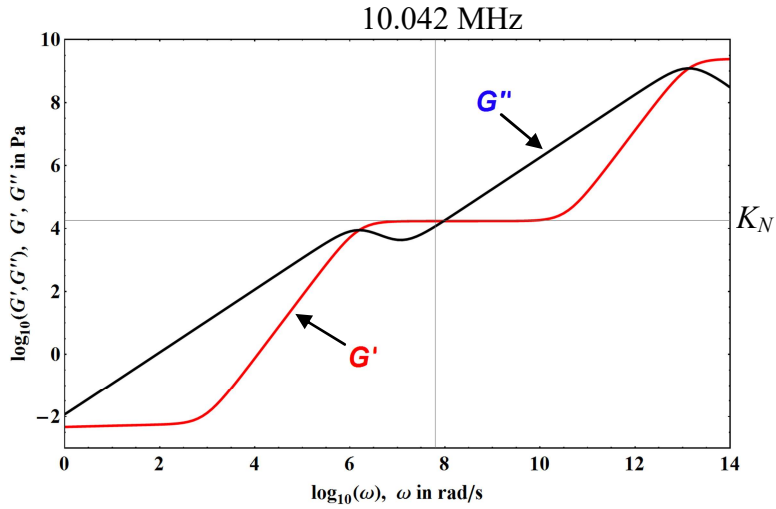

Fig. 16. Logarithmic plot of $G'(\omega)$ (red curve) and $G''(\omega)$ (black curve) calculated with the present modeling in a large frequency domain. The value of $G(t=0)$ is reduced by a factor 0.83. $T = 293.65$ K, $\alpha = 0.04$, $\xi^* = 1$.

We will continue this section by analyzing the pioneering work from Derjaguin *et al.* (Ref. 20) which corresponds to an intermediate frequency of 73.5 kHz from the previous ones. Derjaguin *et al.*'s experiment is quite similar to that of Bund *et al.* except that the liquid is sandwiched between two walls. Derjaguin *et al.* were thus able to vary the thickness $e$ of the liquid layer between 0.9 μm and 3 μm with a total of 5 different thicknesses. They then showed that the values of *G'* and *G''* do not depend on the thickness in their experimental conditions, namely with strains of the order of few percent. However, in appendix B, it will be shown that for larger ranges of variation of the sample thickness, that the invariance found by Derjaguin *et al.* ceases so that *G'* and *G''* finally decrease according to a complex variation law.

The value obtained by Derjaguin *et al.* are *G'* = 31 000 Pa and $\tan(\delta) = 0.3$. More recently Badmaev *et al.* (Ref. 21) repeated the same experiment and found *G'* = 28 000 Pa with the same $\tan(\delta)$. Even more recently Badmaev *et al.* (Ref. 22) repeated the same experiment by playing on the wettability of the contact surface and found *G'* = 25 000 Pa for a hydrophilic surface and 350 Pa for a hydrophobic surface always with the same $\tan(\delta)$. This shows the importance of the hydrophilic or hydrophobic character on the determination of *G'* and is in line with our previous comments. The value of 350 Pa is not used here, but there is still 20% difference between the highest and the lowest value of *G'*. This gap is quite similar to the uncertainty for the data of Bund *et al.* (Ref. 19).

As the parameters of the model are essentially determined by the geometry of the experiment, we can deduce an equivalent cylindrical radius *R* = 0.257 cm of the sample which fixes the value of $d_N$. The value of *d* is fixed with the different thicknesses *e* of the samples. Given the frequency, the value of $\alpha$ does not play a key role and is arbitrarily set at the same value as for Bund *et al*. Since the value of *G'* is independent of thickness in this micrometer range, it can be deduced that $\xi$ must then vary to "compensate" for the thickness variation. Since in these experiments the shear strain amplitude is smaller than in the plate-plate rheometer experiment, $\xi^*$ is expected to have larger values.

Only the results for the largest and smallest thicknesses are shown below. Fig. 17 shows that the value of *G'* can be perfectly determined by the present modeling by considering two different values of $\xi^*$. These $\xi^*$ values are greater than those obtained in the case of the plate-plate rheometer as expected, but above all the numerical values are perfectly consistent with those needed to analyze the dynamic viscosity data of Osipov *et al.* (Ref. 23) in the supercooled phase of water deduced from the flow in a tube having a radius of 1 μm (see Ref. 11). This shows the consistency of the models presented. With these numerical values, it can be deduced that $\xi/e = 51.36\%$ for *e* = 0.9 μm and $\xi/e = 48.80\%$ for *e* = 3 μm. Overall, it appears that the system is highly correlated with an average correlation length of half the thickness. This said, the trend shows that the system is even more solid the thinner the thickness, which is in line with physical intuition.

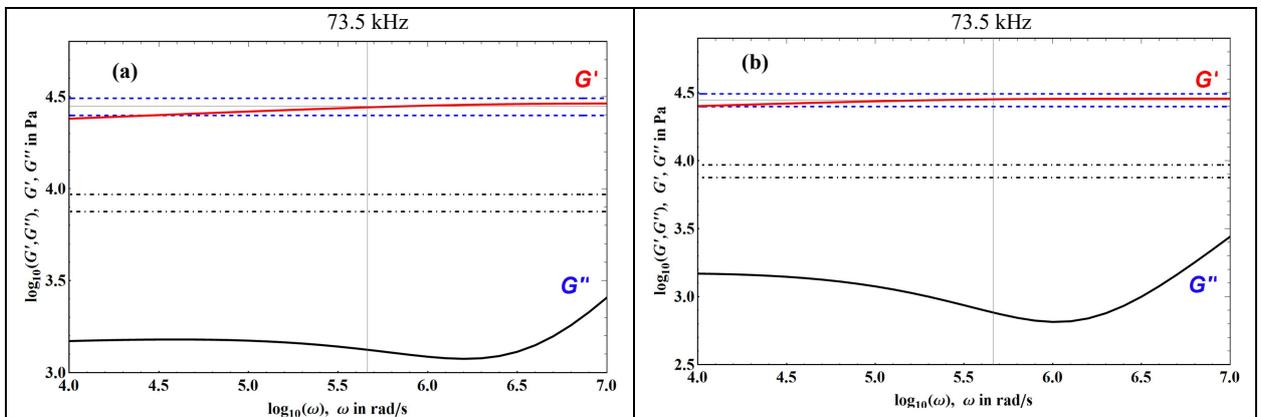

Fig. 17. Logarithmic plot of $G'(\omega)$ (red curve) and $G''(\omega)$ (black curve) calculated with the present modeling in the intermediate frequency range with $\alpha = 0.04$ and *T* = 295.15 K for two thicknesses: (a) *d* = *e* = 0.9 μm and $\xi^*$

= 1350, (b) $d = e = 3$ μm and $\xi^* = 4400$. The blue dashed lines represent the highest and smallest value of $G'$ and the black dot-dashed lines represent the corresponding highest and smallest value of $G''$.

Regarding the value of $G''$, we observe that it is systematically too low compared to the expected value. Since these values are obtained from surfaces with good wettability, the deviation can only be caused by the fluctuations that must exist, as in the case of the plate-plate rheometer. And we saw earlier that the value of $\Delta G''$ due to these fluctuations can be much higher than the theoretical value of $G''$ (i.e. in agreement with the Kramers-Kronig relations). Fig. 18 shows that by adding the same value $\Delta G'' = 7000$ Pa, the present modeling allows to obtain the value of $G''$ deduced from the experimental measurements. This value of $\Delta G''$ may seem high, but relative to the theoretical value of $G''$, the latter is quite comparable, for example, with what has been deduced for $\gamma_0 = 30\%$ in the case of the plate-plate rheometer.

The value of $d_N$ being the same whatever the thickness, so it can be deduced that the value of $K_N$ is a constant independent of the thickness. Fig. 18 shows that the level at $K_N$ exists and is observed at higher frequencies than in Bund *et al.*'s experiment. The value of $K_N$ is higher than the value of $G'$ for the studied frequency as expected. It can be seen in Fig. 18 that the only difference between the two thicknesses is that the level at $K_N$ is narrower in frequency as the thickness decreases, which is related to the fact that $\tau$ also decreases and approaches $\tau_{\mathrm{mi}}$. So at the limit for a quasi-null thickness the present modeling leads to the fact that there are only two levels left for $G'(\omega)$.

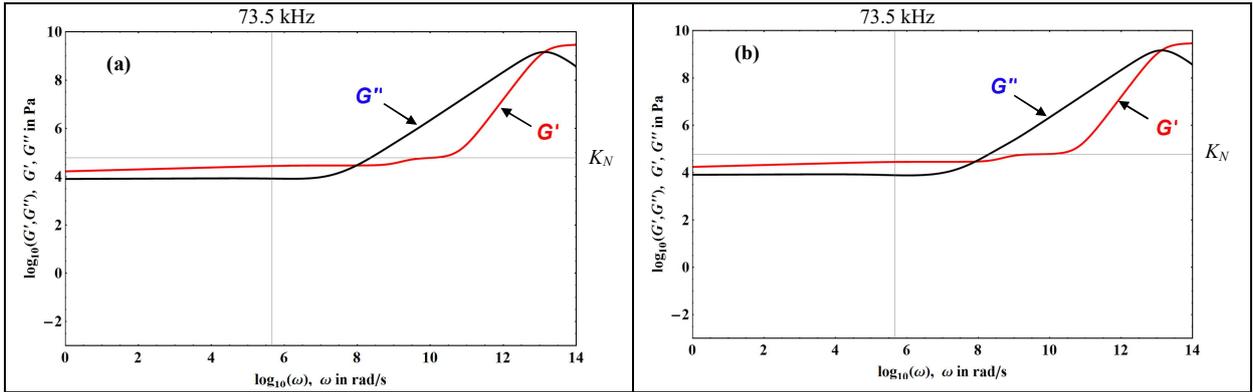

Fig. 18. Modeling in the full frequency range and small gap which shows the logarithmic plot of $G'(\omega)$ (red curve) and $G''(\omega)$ (black curve) versus frequency calculated with the present modeling for which an excess of $G''$ has been added for the two thicknesses: (a) $d = e = 0.9$ μm and $\xi^* = 1350$, (b) $d = e = 3$ μm and $\xi^* = 4400$. In both cases: $T = 295.15$ K, $\alpha = 0.04$ and $\Delta G'' = 7000$ Pa.

To conclude this section, the recent work from Li *et al.* (Ref. 24), which extends the work of Derjaguin *et al.* for much smaller thicknesses ($e \leq 2$ nm) and lower frequencies (between 50 Hz and 2 kHz), will be analyzed. The experiments presented in Ref. 24 were performed using a high-resolution atomic force microscopy measurement (AFM) with silicon tips having radii of $40 \pm 10$ nm. The large radius of curvature of the tips compared to their distance from the opposite mica plate allows to approximate this device as a rheometer with parallel plates separated by a gap distance $e$. A sinusoidal strain is applied to one of the plates at the frequency $\omega$. The strain amplitude is denoted $\gamma_0 = X_0/e$ where $X_0$ is the lateral displacement amplitude of the tip during the oscillations. The data for $G'$ and $G''$ were deduced from a parallel plates geometry modeling: Fig. 2 of Ref. 24 represents the data $G'$ and $G''$ in water as function of the gap distance $e$ for a fixed value of $X_0$ and a fixed frequency of 955.3 Hz while Fig. 3 of Ref. 24 represents the data of $G'$ and $G''$ in water as a function of $\gamma_0$ for a fixed gap distance $e = 0.4$ nm but for three different frequencies. Thus, for the gap distance $e = 0.4$ nm, it is possible to superimpose the data of Fig. 2a'-a'' and 3b'-b'' at the frequency of 955.3 Hz as shown on Fig. 19. It is clear that the only consistent point between the two data sets corresponds to $\gamma_0 = 1$ because it corresponds to the same $X_0 = 0.4$ nm and the same $e = 0.4$ nm; but even so, for this experimental condition, it is observed that $G''$ is shifted between the two data sets by a value larger than the authors' error bars. In other words, the absolute values of $G''$ for $\gamma_0 = 1$ are not very reliable. It follows that the only possible frequency analysis of these data can be performed for $\gamma_0 = 1$ with a gap distance $e = 0.4$ nm.

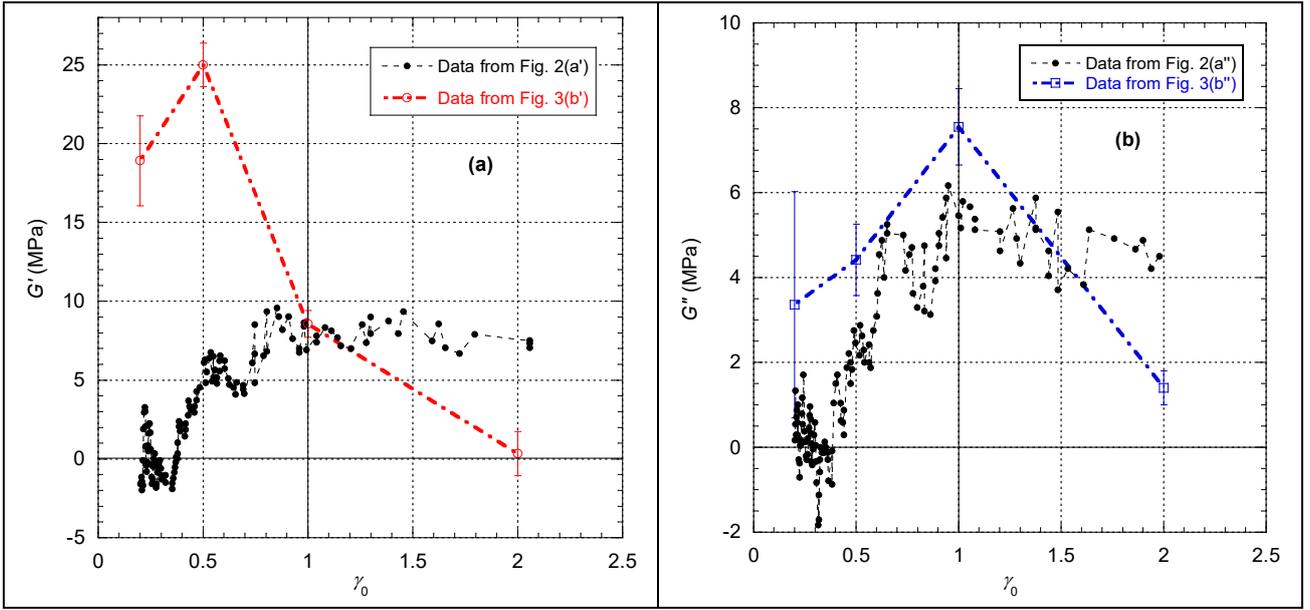

Fig. 19. Plot of experimental shear modulus data for water (from Ref. 24) as a function of $\gamma_0 = X_0/e$ for $e = 0.4$ nm and a fixed frequency of 955.3 Hz at $T = 300$ K. (a) $G'$; (b) $G''$. In both cases, the dashed lines are eyes guides.

To perform the quantitive frequency data analysis for $\gamma_0 = 1$, we need to set the values of our model parameters $d$ and $d_N$. We have shown previously that for a plate-plate rheometer, we have $d = e$. Concerning $d_N$, we have seen that it must be identified with the equivalent radius of the rheometer. Li *et al.* have considered for the calculation of the values of $G'$ and $G''$, a piece of spherical cap of height $\Delta h = 0.25$ nm with respect to the apex as the area $A$ of the upper moving plate. Taking into account the uncertainty on the radius of the spherical cap, the value of the area $A$ can be between 47 nm$^2$ and 78.5 nm$^2$. But the value of $A$ is not given in Ref. 24. We have then arbitrarily fixed the value of $A$ to 75 nm$^2$, conforming to the value reported in Ref. 25. This leads us to fix the value of $d_N$ to 4.886 nm. Fig. 20 shows that the experimental data can be correctly reproduced by setting the parameters $\alpha$ and $\xi^*$ such that: $\alpha = 0.24$ and $\xi^* = 4$.

It can be seen that the value of $\xi^*$ is consistent with those obtained in Fig. 14. Now, the value of $\xi^*$ obtained is interesting because, taking into account the value of $\xi_0$, we deduce that $\xi = 0.51$ nm, thus $\xi$ is slightly larger than $e$. But the value of $e$ given in Ref. 24 corresponds to the distance between the mica plate and the apex of the tip. However, the water rises along the silicone tip by a height $\Delta h = 0.25$ nm. Therefore the value of $d$ should be between $e = 0.4$ nm and $e+\Delta h = 0.65$ nm which corresponds to an average value close to $\xi$. Increasing the gap distance of the equivalent plate-plate rheometer increases the values of $G'$ and $G''$ in the calculations of Li *et al*. Increasing these values does not change the value of $\xi$ but slightly decreases the value of $\alpha$. Given the uncertainty on the value of the area $A$ which also sets the absolute values of $G'$ and $G''$, it is not possible to make a more refined analysis. Despite these uncertainties, an interesting result appears which shows that, for $\gamma_0 \leq 100\%$, the whole sample is correlated, i.e. behaves like a solid, as expected from all the previous data analyses.

Fig. 20b shows that with $\Delta G'' = 2.7$ MPa, the present modeling agrees with the points at 52 Hz and 955.3 Hz. This value of $\Delta G''$ seems large but it is consistent with the very large fluctuations observed in Fig. 2 of Ref. 24. It can also be noticed that the dissipation is stronger than for Derjaguin *et al*. and Badmaev *et al*. experiments with here a $\tan(\delta) \approx 0.62$. The point at 1968.9 Hz, on the other hand, seems inconsistent with the other experimental points. We have seen that a first explanation can be attributed to a shift in the numerical values as can be observed with the points at 955.3 Hz or as possibly to an under estimation of the error bar for this point. Another possibility is a significant reduction in fluctuations for this measurement corresponding to $\Delta G'' = 0.6$ MPa. The data do not allow further analysis so that it can just be deduced that $\Delta G''$ is between the two obtained values. Nevertheless, the agreement between these experimental data and the present modeling is here globally less good than with the previous examples were the measurements were performed with relatively much larger sample volumes and a rather small input strain amplitude of the order of a few percent. On the contrary, here, the strain used is 100%, and probably outside the linear regime, where the output stress signal is no more sinusoidal (e.g. see Fig. 2c of Ref. 13). In this case, in all rigor, the linear approximation can no more be used and the concepts of $G'$ and $G''$ do not make much sense, so their use can only be a rough approximation. Similarly for our model, the replacement of discrete sums by integrals leading to the function $H_N(\nu)$ (see Ref. 11) may not be fully justified when some dimensions of the sample are reduced to a few molecular sizes, as is the case here with thicknesses of the order of nanometer. All these reasons could explain why the analysis of these data is only semi-quantitative.

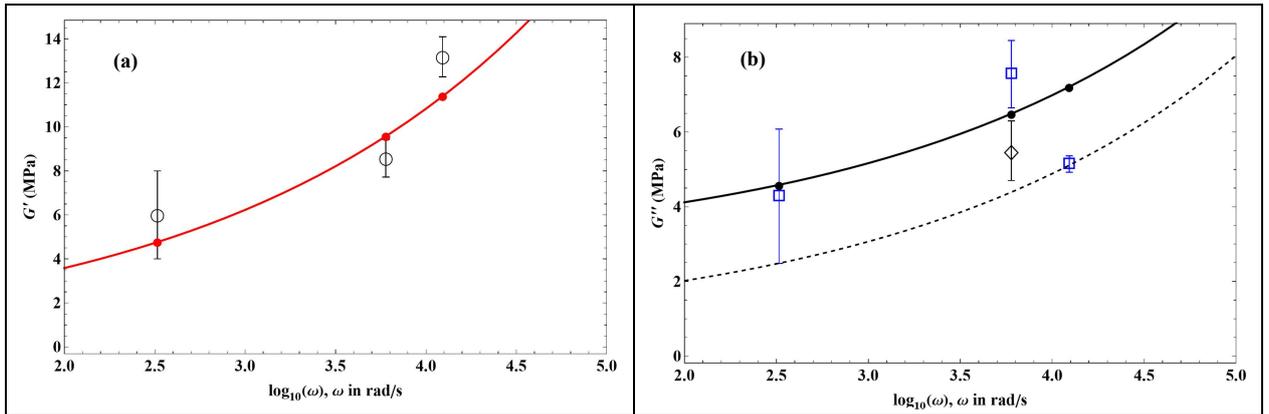

Fig. 20. Semi-logarithmic plot of $G'(\omega)$ and $G''(\omega)$ experimental data for water with $\gamma_0 = 100\%$ (open black circles and open blue squares from Fig. 3 of Ref. 24, and an open black diamond from Fig. 2a'' of Ref. 24) with the present modeling (red curve for $G'$, solid and dashed black curves for $G''$): (a) the red points highlight the theoretical values corresponding to the three experimental frequencies; (b) the solid curve corresponds to $\Delta G'' = 2.7$ MPa with black dots that highlight the theoretical values at the three experimental frequencies and the dashed curve corresponds to $\Delta G'' = 0.6$ MPa. In both cases: $T = 300$ K, $\xi^* = 4$, $\alpha = 0.24$ and $d = e = 0.4$ nm.

Fig. 21 shows the variation of $G'$ and $G''$ over a wide frequency range, encompassing the three experimental frequencies of Li *et al*. Considering the very small value of $d_N$ due to the very small sample volume, it can be seen that the value of $K_N$ is close to the value of the shear elastic constant $K$. As expected from the analysis of Derjaguin's data, it can be seen that the plateau at $K_N$ is very thinly spread out in frequency and occurs at higher frequencies than for the data of Derjaguin *et al*.

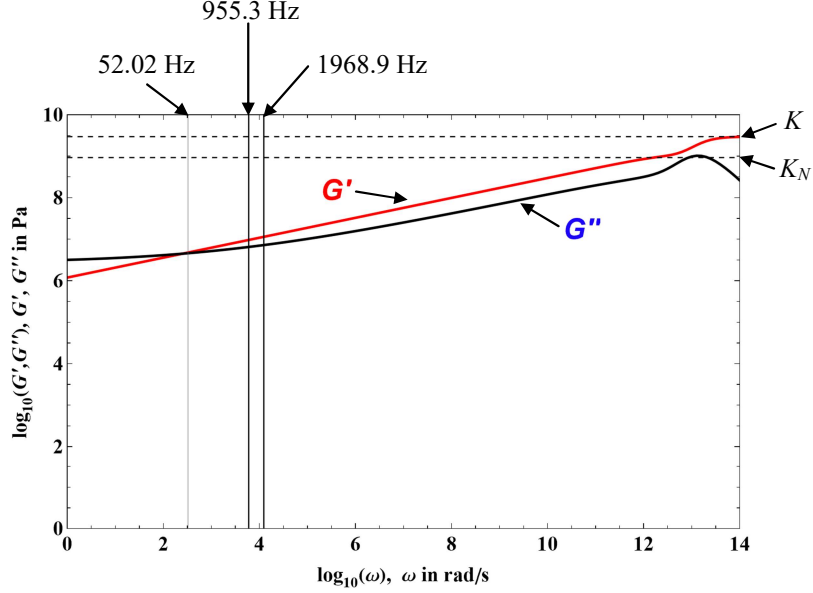

Fig. 21. Modeling in the full frequency range for $\gamma_0 = 100\%$ which shows the logarithmic plot of $G'(\omega)$ (red curve) and $G''(\omega)$ (black curve) versus frequency calculated with the present modeling for which an excess of $G''$ has been added: $T = 300$ K, $\xi^* = 4$, $\alpha = 0.24$, $d = e = 0.4$ nm and $\Delta G'' = 2.7$ MPa.

It can be concluded that all these measurements using different setups are consistent with each other given the very different geometrical parameters and the various frequencies used in these independent experiments.

## 4 Conclusion

In this paper, we first developed a theoretical model to describe the mean trajectory of a basic unit when a fluid system is set out-of-equilibrium upon applying an external mechanical action. This theoretical model has been constructed by analogy with the volume theory of Ref. 11. It takes into account various rheological situations assuming an intrinsic occurrence of a dynamic solid-liquid transition analogous to the thermal ordered-disordered transition of the volume theory.

The laminar Newtonian regime is obtained as the asymptotic limit of the present modeling when the reduced action temperature $T_A^*$ is very large in front of the unit value, i.e. when the local action is very large compared to the global reaction of the system. Within this limit, we have been able to demonstrate the basis of the phenomenological relations introduced in Ref. 11 to describe the dynamic viscosity. This shows that the concept of Newtonian liquid does not contradict the existence of a finite shear elasticity. The existence of a solid-liquid transition in the present modeling implies the existence of a mechanical energy per unit volume (associated with the mechanical action made on the system) so-called *threshold energy*, results in the existence of a *threshold stress* for the fluid whatever the rheological behavior that follows. In other words, a finite stress is always necessary to cause the motion of a fluid which persists over an infinitely long time. This threshold stress is not an intrinsic property of the fluid in the sense that it strongly depends on the nature of the surfaces through the parameter $\xi$ and on the geometry of the experiment through the parameters $d$ and $d_N$.

We then showed that the analysis of the liquid water mechanical relaxation data fits very well with the whole theoretical modeling in the sense that the deduced parameters are consistent with those already determined with the transport coefficients data (Ref. 11). The coherence of this analysis with a more classical analysis of data from measurements imposing an oscillatory shear strain validates the theoretical approach developed.

The whole analysis has shown that the shear moduli $G$, $G'$ and $G''$ are scaled by the shear elastic constant $K$, but are not constants of the medium in the sense that their values depend on the experimental conditions. Furthermore, it has been shown that the fluctuations of $G$ play a determining role in the value of $G''$ and explain why the Kramers-Kronig relations apparently seem to be violated.

It is concluded that the transport properties as well as the viscoelastic properties of fluids can be consistently modeled and analyzed assuming the existence of finite static shear elasticity in liquids. The fact that any finite volume of liquid at complete mechanical rest, or subjected to a mechanical stress of sufficiently small amplitude, must be considered as a solid, implies the possibility of observing new physical phenomena specific to solids such as thermoelasticity. A thermoelastic conversion of the mechanical energy has been already identified in response to a shear strain field or under microflow in liquid water, glycerol, being more pronounced for polymer melts (e.g. Refs. 26-27).

This new approach opens perspectives to explore and offer a better understanding of different problems such as Rayleigh-Bénard type flow instabilities, wetting phenomena or microfluidic flow which is typically the scale at which the physiological exchanges (e.g. Ref. 28) and fluid transport take place.

## 5 ACKNOWLEDGEMENTS

We thank Ing. P. Baroni and Dr. L. Noirez from the Léon Brillouin laboratory (CEA, France) for providing their experimental data and to Dr. L. Noirez for many helpful suggestions and discussions.

This work benefited from the support of the project ZEROUATE under Grant ANR-19-CE24-0013 operated by the French National Research Agency (ANR).

## 6 APPENDIX A: Mechanical energy functional derivation

It has already been shown in the case of the spatial model (Ref. 11) that the introduction of fractional derivatives of order $1+v/2$ in the excess elastic energy functional is mathematically equivalent to the replacement of an individual elastic constant $K_q$ by a power law $K_q = K\left(\dfrac{q}{q_c}\right)^v$ in the Fourier space.

Since the fundamental assumption of the present modeling is also the existence of a mechanical energy functional $F_A$ (given by Eq. (1) for translational motion and by Eq. (2) for rotational motion), we are going to demonstrate that the introduction of fractional derivatives in $F_A$ is mathematically equivalent to the replacement of $K_A^{t,r}$ by a power law such as Eq. (13) in the Fourier space.

We will demonstrate only for the translation but we can do the same by analogy for the rotation. The starting point is to replace Eq. (1) with the following fractional derivative relation:

$$F_A^t = c_0 \, \omega_c^2 \int\limits_{\substack{\text{duration} \\ \text{of action}}} K_{A0}^t \lim_{t_2 \to t_1} \left( \frac{\partial^{\nu_A} x_f}{\partial (\omega_c t_1)^{\frac{\nu_A}{2}} \partial (\omega_c t_2)^{\frac{\nu_A}{2}}} \right) dt_1 \tag{A.1}$$

Since the function $x_f(t_1, t_2)$ is expanded in Fourier series, Eq. (A.1) involves only fractional derivatives of $\exp(i\omega t)$ and we have already shown in appendix A of Ref. 11 that it leads to a complex exponential function.

Now inserting Eq. (5) into Eq. (A.1), and using the orthogonality property of imaginary exponential functions, one obtains the following expression for $F_A^t$:

$$F_A^t = c_0 \, t \sum_{\omega = \frac{\omega_c}{N_A}}^{\omega_c} \left( K_{A0}^t \, \omega_c^2 \left| \frac{\omega}{\omega_c} \right|^{\nu_A} x_{f,\omega} \right) \tag{A.2}$$

Finally, it appears that the introduction of fractional derivatives in (A.1) is mathematically equivalent to the replacement of $K_A^{t,r}$ by Eq. (13) in the mechanical energy functional expression (i.e. in Eq. (9) or Eq. (10)).

## 7 APPENDIX B: Elastic shear moduli versus the sample liquid thickness

All the data analyzed up to this point in the paper did not concern the dependence of $G'$ and $G''$ on the sample liquid thickness $e$, except in the case of Derjaguin *et al.* (Ref. 20). However, such data of $G'$ and $G''$ versus $e$ exist for very different substances in the literature and recent works (e.g. Refs. 10a and 29) mention that the power law $G'(e) \approx \beta \, e^{-3}$, with $\beta \sim$ 1 nJ, agrees with these data. For example Zaccone *et al.* wrote in Ref. 29:

"[…] this law appears to be truly universal, and we present theoretical fittings of several, very different systems in Figure 3."

It will be shown in this appendix that this law is only a rough approximation which diverges when the thickness $e$ tends to zero contrary to the experimental data such as for example those of Derjaguin *et al.* (Ref. 20) where an invariance of $G'$ and $G''$ at micrometer thicknesses is found. The same authors seem to have taken the measure of the incompatibility of their law at very low thicknesses since they wrote in Ref. 29 about experimental data on the o-Terphenyl (OTP):

"Those data also show the $\sim L^{-3}$ [i.e. $\sim e^{-3}$ with the present notations] behavior, but the data at the shortest confinement length, $\sim 0.01$ mm, suggest the possible existence of a plateau upon going toward lower $L$, while the experimental accuracy is lowered as the confinement increases. The paucity of experimental data does not allow for drawing a definitive conclusion on this effect (i.e., the possible existence of a plateau in $G'$ at low $L$ in certain systems), which should also be the object of further investigation, both experimentally and in theory."

Regardless of the uncertainty in the data, it is more accurate to say that most experimental data show a plateau at low thicknesses and therefore there is a more general and complex law $G'(e)$ which accounts for the experimental data in all their ranges of variation and which leads to a finite value of $G'$ when the thickness tends to zero.

The general expression of the law $G'(e)$ is given here in an empirical form, but it is related in all rigor to the evolution of the parameter $\overset{\leftrightarrow}{\xi}$ of the present modeling. Thus, the variation law $G'(e)$ can be written in the following general form:

$$G'(e) = G'(0) \left\{ \frac{a}{\left( 1 + \dfrac{1}{3}\left( \dfrac{e}{l_1} \right)^{\alpha_1} \right)^3} + \frac{1-a}{\left( 1 + \dfrac{1}{3}\left( \dfrac{e}{l_2} \right)^{\alpha_2} \right)^3} \right\} \tag{B-1}$$

where $a$, $l_1$, $l_2$, $\alpha_1$ and $\alpha_2$ are empirically determined constants. Eq. (B-1) has several features: it leads to a finite value of $G'$ when $e$ tends to 0, i.e. $G'(0)$, with a horizontal tangent as suggested by the data of Derjaguin *et al*. It also allows to account for the abrupt decrease of the data around a given thickness through the first term in the square brackets. The two terms between the square brackets in Eq. (B-1) represent generalized exponentials similar, for example, to the long-time relaxation term used in Eq. (33).

We will begin by comparing Eq. (B-1) with the water data of Li *et al*. (Ref. 24). Fig. 22 shows, first of all, that the experimental data $G'$ and $G''$ of water for the smallest value of displacement $X_0 = 0.4$ nm display a flattening of the variations at the smallest thicknesses. Then it is observed that Eq. (B-1) is able to account for the data of $G'$ as well as $G''$, taking into account the experimental fluctuations.

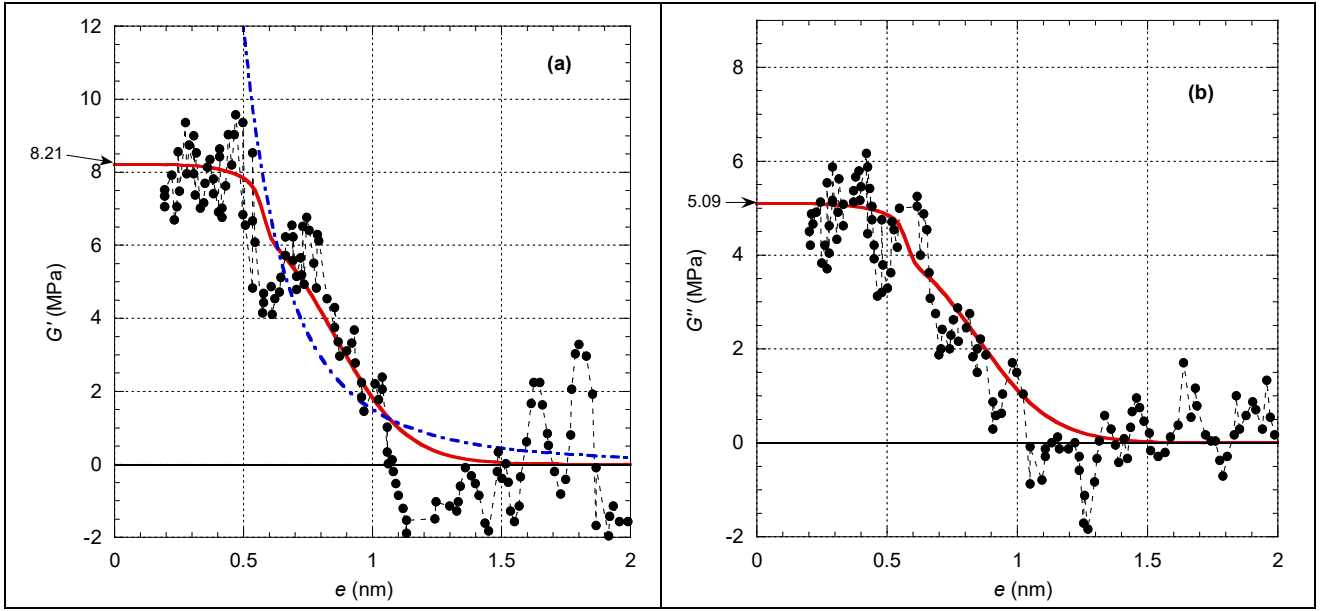

Fig. 22. Plot of $G'$ and $G''$ experimental data (black dots link by a dashed line) for water (from Fig. 2a'-a'' of Ref. 24) as a function of the sample liquid thickness $e$ at $T = 300$K. In both cases, the red curves correspond to Eq. (B-1) with $a = 0.15$, $l_1 = 0.58$ nm, $l_2 = 0.9$ nm, $\alpha_1 = 30$ and $\alpha_2 = 5$. (a) The blue dot-dashed curve represents the power law $G'(e) = 1.5/e^3$; (b) the red curve is simply determined as $G''(e) = \tan(\delta) \times G'(e)$ with $\tan(\delta) = 0.62$.

As the authors of Ref. 29 suggested, it is interesting to analyze the data on OTP which corresponds to a small molecule as water. Fig. 23 shows that the experimental data on OTP are very comparable to those of water with the appearance of a very clear plateau at low thicknesses, a plateau that was also pointed out by the authors of the data in Ref. 30. It is observed that Eq. (B-1) allows us to reproduce these data correctly with the same exponents as for water while the power law can only approximately satisfy at the three points corresponding to the largest thicknesses.

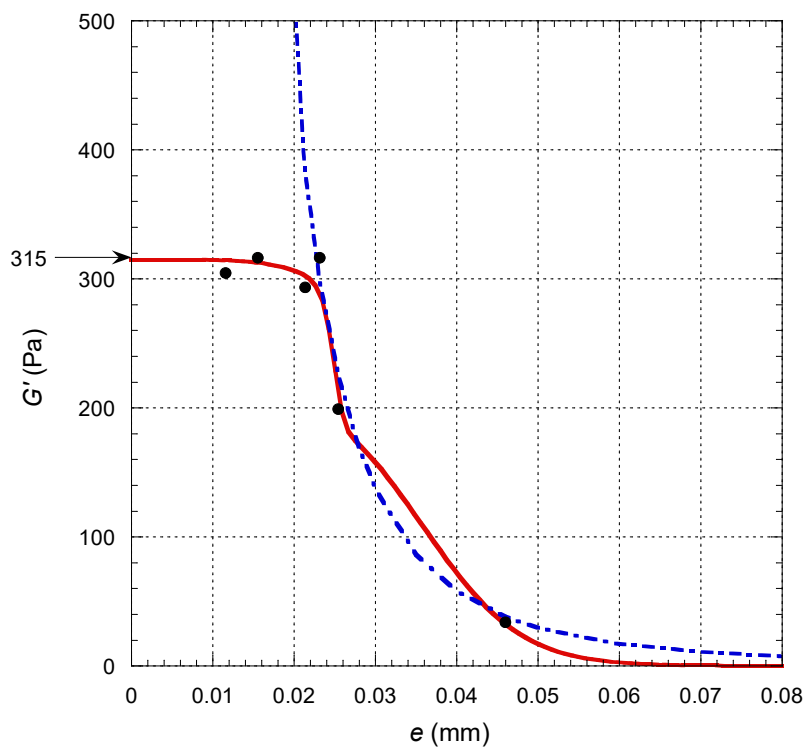

Fig. 23. Plot of $G'$ experimental data (black points) for OTP (from Fig. 3a of Ref. 30) as a function of the sample liquid thickness $e$ at $T = 339.15$ K. The red curves correspond to Eq. (B-1) with $a = 0.33$, $l_1 = 0.025$ mm, $l_2 = 0.038$ mm, $\alpha_1 = 30$ and $\alpha_2 = 5$. The blue dot-dashed curve represents the power law $G'(e) = 0.0037/e^3$.

In Fig. 24, we show the analysis of experimental data for three polymer systems that are more complex than water and OTP. Fig. 24a shows a flattening of the data for a polybutylacrylate (PBuA) at low thicknesses while in Fig. 24b, for a methoxy-phenyl benzoate substituted polyacrylate (PAOCH3), it is observed a plateau as for water and OTP. These effects were moreover highlighted by the authors respectively in Refs. 30 and 31. Fig. 24a shows very clearly the break that could be guessed for water and OTP and it can be seen that Eq. (B-1) accounts well for this break. On this same figure, it is observed that only a part of the data at high thicknesses can be approximated by a power law but whose exponent is different from the value 3. The same kind of comment applies to the other two figures. Concerning Fig. 24c, the authors have mentioned in the legend of their Fig. 9 (Ref. 32) that those can be fitted with a power law whose exponent is of the order of 2.5.

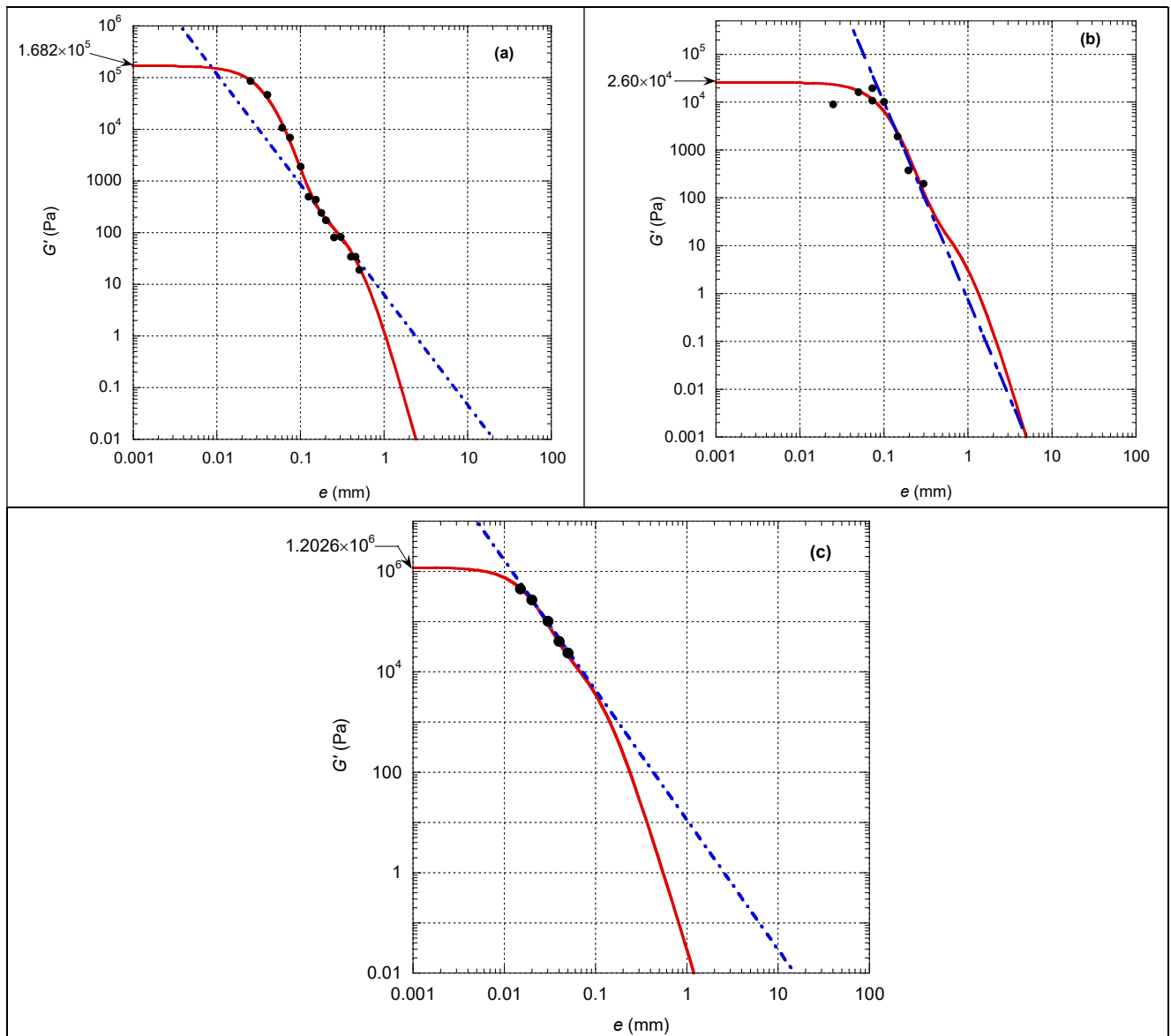

Fig. 24. Logarithmic plot of $G'$ experimental data (black points) for three different substances versus the sample liquid thickness $e$. (a) Experimental data for PBuA at $T = 298.15$ K from Fig. 4b of Ref. 30; the red curve corresponds to Eq. (B-1) with $a = 0.99856$, $l_1 = 0.0298$ mm, $l_2 = 0.2623$ mm, $\alpha_1 = \alpha_2 = 2$ and the blue dot-dashed curve represents the power law $G'(e) = 6.31/e^{2.134}$ . (b) Experimental data for PAOCH3 at $T = 403.15$ K from an insert in Fig. 1c of Ref. 31; the red curve corresponds to Eq. (B-1) with $a = 0.99856$, $l_1 = 0.075$ mm, $l_2 = 0.5$ mm, $\alpha_1 = \alpha_2 = 2$ and the blue dot-dashed curve represents the power law $G'(e) = 0.741/e^{4.127}$ . (c) Experimental data for polystyrene from Fig. 9 of Ref. 32; the red curve corresponds to Eq. (B-1) with $a = 0.98$, $l_1 = 0.0139$ mm, $l_2 = 0.06$ mm, $\alpha_1 = \alpha_2 = 2$ and the blue dot-dashed curve represents the power law $G'(e) = 11.316/e^{2.583}$ .

In conclusion of this appendix, it can be said that the experimental data do not agree well with a power law with exponent value equal to 3. However, in certain well-chosen thickness ranges, a power law with an exponent smaller or larger than 3 can agree with the data. On the other hand, the empirical form of Eq. (B-1) is able to account for all the experimental data. This equation has the advantage of being able to determine a value for $G'(0)$ that relates to the value of $K_N$ in the theoretical approach presented here. As a last remark, we stress that Eq. (B-1) is quite general since it applies to both small molecules and polymers. Besides the very different values of $G'(0)$, which is not surprising, the difference between the two types of substances lies in the value of the exponents $\alpha_1$ and $\alpha_2$ which determine the rate of change with thickness. For the former, which are pure substances made up of a single type of molecule, the transition between the two behaviors is strong, with exponents, $\alpha_1 = 30$ and $\alpha_2 = 5$, whereas for polymers, it is much softer, with exponents $\alpha_1 = 2$ and $\alpha_2 = 2$. It is likely that at least some of the change in this behavior is associated with polydispersity, which logically has the effect of broadening the transitions. This could be tested by making measurements with mixtures of small molecules.

## 8 REFERENCES

[1] W.M. Slie, A.R. Donfor, and T.A. Litovitz, J. Chem. Phys. **44**, 3712 (1966).

[2] B.V. Derjaguin, U.B. Bazaron, K.T. Zandanova, and O.R. Budaev, Polymer. **30**, 97 (1989).

[3] L. Noirez, P. Baroni, H. Mendil-Jakani, Polymer International **58**, 962–968 (2009).

[4] L. Noirez and P. Baroni, J. Mol. Struct. **972**, 16 (2010).

[5] Y. Chushkin, C. Caronna, and A. Madsen, EPL (2008).

[6] D.P. Shelton, J. Chem. Phys. (2014).

[7] J. Frenkel, *Kinetic Theory of Liquids*, Oxford University Press, Oxford (1946).

[8] R. Zwanzig, R. D. Mountain, J. Chem. Phys. **43**, 4464 (1965).

[9] J. D. Ferry, *Viscoelastic Properties of Polymers*, Wiley, 3rd Edition (1980).

[10] (a) A. Zaccone, K. Trachenko, PNAS **114**, 19653 (2020); (b) M. Baggioli, M. Vasin, V. V. Brazhkin, and K. Trachenko, Phys. Rev. D **102**, 025012 (2020); (c) M. Baggioli, M. Vasin, V. Brazhkin, and K. Trachenko, Phys. Rep. **865**, 1–44 (2020); (d) M. Baggioli, M. Landry, and A. Zaccone, Phys. Rev. E **105**, 024602 (2022).

[11] F. Aitken and F. Volino, Phys. Fluids **33**, 117112 (2021).

[12] F. Aitken and F. Volino, Phys. Fluids **34**, 017112 (2022).

[13] L. Noirez, P. Baroni, and H. Cao, J. Mol. Liq. **176**, 71 (2012).

[14] L. Noirez and P. Baroni, J. Phys. Condens. Matter **24** (2012).

[15] P. Baroni, H. Mendil and L. Noirez, 2005 Fr. Pat. 05 10988.

[16] W. Wagner and A. Pruß, J. Phys. Chem. Ref. Data **31**, 387 (2002).

[17] M.L. Huber, R.A. Perkins, A. Laesecke, D.G. Friend, J.V. Sengers, M.J. Assael, I.N. Metaxa, E. Vogel, R. Mareš, and K. Miyagawa, J. Phys. Chem. Ref. Data **38**, 101 (2009).

[18] Y. Xie, K.F. Ludwig, G. Morales, D.E. Hare, and C.M. Sorensen, Phys. Rev. Lett. **71**, 2050 (1993).

[19] A. Bund and G. Schwitzgebel, Anal. Chem. **70**, 2584 (1998).

[20] B.V. Derjaguin, U.B. Bazaron, K.D. Lamazhapova, and B.D. Tsidypov, Phys. Rev. A **42**, 2255 (1990).

[21] B.B. Badmaev, O.R. Budaev, T.S. Dembelova, and B.B. Damdinov, Ultrasonics **44**, 1491 (2006).

[22] B. Badmaev, T. Dembelova, B. Damdinov, D. Makarova, and O. Budaev, Colloids Surfaces A Physicochem. Eng. Asp. **383**, 90 (2011).

[23] Y. Osipov, B. Zheleznyi, and N. Bondarenko, Russ. J. Phys. Chem. **51**, 748 (1977).

[24] T. De Li, & E. Riedo, Phys. Rev. Lett. **100**, 106102 (2008).

[25] T. De Li, J. Gao, R. Szoszkiewicz, U. Landman, and E. Riedo, Phys. Rev. B - Condens. Matter Mater. Phys. **75**, 1–6 (2007).

[26] E. Kume, P. Baroni and L. Noirez, Scie. Reports **10**, 13340 (2020).

[27] L. Noirez, P. Bouchet and P. Baroni, J. Phys. Chem. Lett. **4**, 2026−2029 (2013).

[28] U. Windberger, P. Baroni and L. Noirez, J. Biomed Mater Res. Part A, 110(2), 298– 303 (2022).

[29]  A. Zaccone and L. Noirez, J. Phys. Chem. Lett. **12**, 650–657 (2021).

[30]  L. Noirez, H. Mendil-Jakani and P. Baroni, Philos. Mag. **91**, 1977–1986 (2011).

[31]  H. Mendil, P. Baroni and L. Noirez, Eur. Phys. J. E **19**, 77–85 (2006).

[32]  D. Collin and P. Martinoty, Phys. A Stat. Mech. its Appl. **320**, 235–248 (2003).